\documentclass[journal]{IEEEtran}

\ifCLASSOPTIONcompsoc

\usepackage[nocompress]{cite}
\else
  \usepackage{cite}
\fi

\usepackage{amsmath,amsfonts,amssymb}
\usepackage{bm} 
\usepackage{cases}
\usepackage{float}
\usepackage{mwe}
\usepackage{mathtools}
\usepackage{comment}
\usepackage{empheq}
\usepackage{booktabs}
\usepackage{textcomp}
\usepackage{algorithm,algorithmic}
\usepackage{array}
\usepackage{subfig}
\usepackage{url}
\usepackage{caption}
\usepackage{multirow}
\usepackage{systeme}
\usepackage{soul}

\newtheorem{theorem}{Theorem}

\usepackage{tikz}
\usetikzlibrary{positioning}
\usetikzlibrary{3d}
\usetikzlibrary{matrix}
\usetikzlibrary{decorations.text}
\usetikzlibrary{spy}

\makeatletter
\tikzoption{canvas is plane}[]{\@setOxy#1}
\def\@setOxy O(#1,#2,#3)x(#4,#5,#6)y(#7,#8,#9)%
{\def\tikz@plane@origin{\pgfpointxyz{#1}{#2}{#3}}%
	\def\tikz@plane@x{\pgfpointxyz{#4+#1}{#5+#2}{#6+#3}}%
	\def\tikz@plane@y{\pgfpointxyz{#7+#1}{#8+#2}{#9+#3}}%
	\tikz@canvas@is@plane
}
\makeatother
\tikzoption{canvas is xy plane at z}[]{%
	\def\tikz@plane@origin{\pgfpointxyz{0}{0}{#1}}%
	\def\tikz@plane@x{\pgfpointxyz{1}{0}{#1}}%
	\def\tikz@plane@y{\pgfpointxyz{0}{1}{#1}}%
	\tikz@canvas@is@plane
}
\makeatother

\usepackage[english]{babel}
\usepackage[utf8]{inputenc} 
\usepackage[bookmarks=false,hyperfootnotes=false]{hyperref}
\hypersetup{
			colorlinks=true,
			linkcolor=blue,
			anchorcolor=black,
			citecolor=green,
			urlcolor=blue
			}
\usepackage{tipa}
 \expandafter\let\csname equation*\endcsname\relax
 \expandafter\let\csname endequation*\endcsname\relax

\newcommand{\RR}{\mathbb{R}}

\newcommand{\Abold}{\mathbf{A}}

\newcommand{\Dbold}{\mathbf{D}}

\newcommand{\Hbold}{\mathbf{H}}

\newcommand{\Mbold}{\mathbf{M}}

\newcommand{\Sbold}{\mathbf{S}}

 \newcommand{\fbold}{\mathbf{f}}
 \newcommand{\gbold}{\mathbf{g}}

 \newcommand{\sbold}{\mathbf{s}}
 \newcommand{\tbold}{\mathbf{t}}
 \newcommand{\ubold}{\mathbf{u}}

\newcommand{\zbold}{\mathbf{z}}

\newcommand{\lambdabold}{\bm{\lambda}}

\DeclareMathOperator*{\argmin}{arg\,min}

\let\labelindent\relax 
\usepackage{enumitem}

\DeclareUnicodeCharacter{2212}{-}

\begin{document}
%

\title{Efficient $\ell^0$ gradient-based Super Resolution for simplified image segmentation}
%
%
%

\author{Pasquale~Cascarano,~Luca Calatroni,~Elena~Loli Piccolomini 
\thanks{P. Cascarano is with the Department of Mathematics, University of Bologna, 40126, Bologna, Italy. Correspondence e-mail: \href{mailto:pasquale.cascarano2@unibo.it}{pasquale.cascarano2@unibo.it}}
\thanks{L. Calatroni is with CNRS, Université C\^{o}te d'Azur, INRIA, I3S, UMR 7271, Sophia-Antipolis, France.}
\thanks{E. Loli Piccolomini is with the Department of Computer Science and Engineering, University of Bologna,  40126, Bologna, Italy.}
}

%
%

\maketitle

\begin{abstract}
We consider a variational model for single-image super-resolution  based on the assumption that the gradient of the target image is sparse.  We enforce this assumption by considering both an isotropic and an anisotropic $\ell^0$ regularisation on the image gradient combined with a quadratic data fidelity, similarly as studied in \cite{Storath14} for general signal recovery problems. For the numerical realisation of the model, we propose a novel efficient ADMM  splitting algorithm whose substeps solutions are computed efficiently by means of hard-thresholding and standard conjugate-gradient solvers. We  test our model on highly-degraded synthetic and real-world data and quantitatively compare our results with several variational approaches as well as with state-of-the-art deep-learning techniques. Our experiments show that $\ell^0$ gradient-regularised super-resolved images  can be effectively used to improve the accuracy of standard segmentation algorithms when applied to QR and cell detection, and landcover classification problems, in comparison to the results achieved by other approaches.
\end{abstract}

\begin{IEEEkeywords}
Single-image super-resolution, $\ell^0$-gradient regularization, inverse Pott function super-resolution,  ADMM.
\end{IEEEkeywords}

%

\IEEEpeerreviewmaketitle

\section{Introduction}

The task of single image Super-Resolution (SR) consists in improving the spatial resolution of an observed Low-Resolution (LR) imaging data so as to obtain a High-Resolution (HR) version which, typically, can be used as a reference for subsequent analysis. Image resolution is limited in many applications due to the optical characteristics and the physical limitations of the acquisition devices. Some standard examples are biomedical and astronomic imaging where, due to light aberration phenomena, close objects (molecules, stars\ldots) on LR images cannot be correctly distinguished/detected, see, e.g.~\cite{Galbraith2011,Willett2004}.
SR techniques are often employed also in image recognition problems. This is the case, for instance, of QR code recognition where images are often captured by scanning tools (e.g. cell-phones) from relatively large distances which may affect the accuracy of the recognition \cite{Kato2011}.
Analogously, in remote sensing applications such as land-cover classification, only LR measurements are often available, which may limit significantly the classification precision \cite{Ling2016,wickham2014}. 


Mathematically, the task can be formulated as an ill-posed inverse problem: for a given vectorised LR image $\gbold \in \mathbb{R}^{M}$, we look for its HR version $\ubold \in \mathbb{R}^{N}$ defined on a space of dimension $N=L^{2}M$ with magnification factor $L >1$ which satisfies the following linear degradation model: 
\begin{equation}\label{eq:model}
    \gbold = \Sbold \Hbold \ubold + \boldsymbol{\eta}.
\end{equation}
Here, $\Sbold \in \mathbb{R}^{M \times N}$ stands for the down-sampling operator,  $\Hbold \in \mathbb{R}^{N \times N}$ describes blur degradation and $\boldsymbol{\eta}$ denotes the realisation of an Additive White Gaussian Noise (AWGN) r.v. with zero mean and standard deviation $\sigma_{\mathbf{\eta}}>0$.

Due to the ill-posedness of the operator $\Sbold\Hbold$, a standard approach for solving \eqref{eq:model} consists in encoding prior knowledge about the solution $\ubold$ and on the data statistics via an energy minimisation approach, so that an approximated solution $\ubold^{*}\in\mathbb{R}^{N}$ is computed by solving
\begin{equation}  \label{eq:reg_mod}
    \ubold^{*} \in  \argmin_{\ubold \in \mathbb{R}^{N}} \dfrac{1}{2} \lVert \Sbold \Hbold \ubold - \gbold \rVert^{2}_{2} + \mu R(\ubold),
\end{equation}
where the quadratic fidelity term models the presence of AWGN while the (possibly non-convex) regularisation term $R: \mathbb{R}^{N}\to \mathbb{R} \cup \left\{+\infty\right\}$ encodes prior information on the target image $\ubold$, thus ensuring the stability of the inversion process. The two terms are balanced by a regularisation parameter $\mu>0$. We refer the reader to \cite{Chaudhuri2001,Yue16} and to the references therein for a review on variational approaches for SR problems.

In this work, we choose $R$ so as to promote gradient sparsity, which is often desirable in image segmentation applications whenever a simplified, edge-preserving version of the original data $\gbold$ is required for further analysis.
In recent years, sparse and non-convex gradient-based regularisation approaches have become very popular in the context of image reconstruction due to their better ability of preserving sharp edges even in low-contrast scenarios. A significant contribution has been made by Storath et al. in a series of papers \cite{Storath14,Storath15,Storath17} where sparsity on the image gradient $\mathbf{D} \ubold\in\mathbb{R}^{2N}$ is promoted by $\ell^0$ regularisation which reads
\begin{equation}
\lVert \mathbf{D} \ubold\rVert_0:=\# \left\{ (\mathbf{D} \ubold)_i, i=1,\ldots 2N: (\mathbf{D} \ubold)_i\neq 0 \right\}.
\end{equation}
This choice has been thoroughly studied for several imaging problems such as deconvolution, sparse recovery, joint reconstruction and segmentation, see, e.g., \cite{Xu11}. Moreover, it has has been shown to be very useful in many situations where a further classification/labelling step is required. In this work we propose a novel numerical scheme endowed with convergence guarantees which justifies the use of this type of regularisation in the context of SR problems with high blur and noise degradation.


\subsection{Related work}
The vast majority of sparse optimisation approaches for SR problems enforces sparsity either on the signal itself \cite{Candes} or its representation w.r.t. to some basis/overcomplete dictionary \cite{Yang2010}. These methods and their non-convex extensions have been shown to be very powerful in several applications such as image microscopy \cite{Gazagnes} where signal-sparsity can be assumed. However, for non-point-like objects (such as piece-wise constant regions), this type of regularisation is not the appropriate choice. 
Other classical approaches to the SR problem are based on the use of least squares, Fourier series and Tikhonov-type gradient regularisations \cite{Chaudhuri2001}, which favour noise removal at the price of creating smoothing and ringing artefacts which are undesirable in many applications such as object detection, where images with sharper edges are preferable for better classification.
To overcome this drawback, the use of edge-preserving convex regularisations based on the idea of gradient sparsity, such as Total Variation (TV) \cite{Chan2007,MarquinaOsher2008,tao2009alternating,Gao2016}, its fractional  \cite{Yao2020} and  $\ell^0$ extension \cite{Storath14,Storath15,Storath17}, has been proposed. Such methods have shown good performances in many applications, although their convexity (in the case of TV) or their challenging numerical realisation (in the case of non-convex approaches) often limit their practical use and precision. Different approaches for solving the SR problem make use of deep architectures encoding prior information on the desired HR solution from a training set of examples \cite{yang2019,wang2020,sun2020learned}. In particular, in \cite{zhang2017} the authors present a Plug-and-Play (PnP) framework \cite{venkatakrishnan2013} which exploits deep convolutional neural network denoisers embedded in standard optimisation algorithms, such as Alternating Direction of Multipliers (ADMM) or Half-Quadratic Splitting (HQS). Differently from model-based variational approaches, deep learning-based methods do not require an explicit expression of the regularisation term $R$, since this can be learned directly from the data and adapted to the particular application considered. Those methods have currently reached state-of-the-art performances in several image reconstruction problems, although their theoretical foundation and their stability to noise perturbations still limits their practical use in the case of highly-degraded image data. 


\subsection{Contribution}

We consider a variational model for solving problem \eqref{eq:reg_mod} where a quadratic data fidelity is combined with an $\ell^0$-gradient regularisation term both in a coupled (isotropic) and decoupled (anisotropic) form, the latter being better suited for directionally-biased images, such as QR scans.
To solve the model efficiently, we propose to use an ADMM algorithm which decomposes the original problem into substeps cheaply solved by means of direct hard-thresholding and standard iterative Conjugate Gradient (CG) linear solvers. Our variable splitting differs from the one introduced by  Storath et al. in \cite{Storath14,Storath15,Storath17}, where the non-convex substeps are solved by means either of approximate graph-cuts approaches \cite{Boykov2001} or 
dynamic programming algorithms. As well as for these different numerical schemes, we prove in this paper fixed-point convergence for the proposed ADMM algorithm.
Up to our knowledge, the same variable splitting has been used only in the case of convex regularisation functions, such as TV, in \cite{tao2009alternating,Gao2016} where convergence to the global minimum is proved.

We test our SR model on real-world applications (QR scanning, cell detection and land-cover labelling) where a simplified HR version of the given LR image $\gbold$ is required in view of further analysis, showing that the proposed model improves significantly 
segmentation and labelling precision. 


\subsection{Organisation of the paper} 

In Section \ref{sec:SRmodel} we provide a review of gradient-sparse variational methods for single-image SR. In Section \ref{sec:ADMM} we present a novel converging ADMM scheme for solving the proposed model along with details on its practical realisation. In Section \ref{sec:modelling} we report some numerical tests on model parameter sensitivity performed on synthetic data. Finally, in Section \ref{sec:experiments} we apply our model to some real-world applications such as QR scanning, cell detection, compressed JPG SR and land-cover classification.
We report the convergence proofs of the proposed ADMM schemes in Appendix \ref{sec:convergence} to improve the flow of the manuscript.

\section{$\ell^0$ gradient-based super-resolution}  \label{sec:SRmodel}

The use of convex gradient-based regularisations for SR problems dates back to \cite{Chan2007,MarquinaOsher2008}, where TV regularisation\footnote{By $\lVert \cdot\rVert$ we denote the standard Euclidean modulus.
 \begin{equation}   \label{eq:TV_reg}
 \lVert \Dbold \ubold \rVert_{1,p} = \sum_{i=1}^{N} \left( \lVert (\mathbf{D_h} \ubold)_i\rVert^p + \lVert(\mathbf{D_v} \ubold)_i\rVert^p\right)^{1/p},
\end{equation}}
was employed to promote sparsity on the image gradient $\Dbold \ubold = (\mathbf{D_h} \ubold, \mathbf{D_v} \ubold)\in \mathbb{R}^{2 \times N} $. 
Note, that for  $p\in\left\{1,2\right\}$ anisotropic/isotropic regularisation is promoted, respectively. We remark that fractional generalisations to exponents $1<p<2$ are also possible \cite{Yao2020}.


Gradient-sparsity can be enforced more severely by means of non-convex $\ell^0$ gradient smoothing, see, e.g., \cite{Xu11} and \cite{Storath14}. Using an analogous notation as in \eqref{eq:TV_reg}, for $p\in\left\{1,2\right\}$ we thus consider the $\ell^0$ gradient regularisation functional defined by:
\begin{align} \label{eq:zero_norm}
R(\ubold)& =\lVert \Dbold \ubold \rVert_{0,p}   \\
& := \sum_{i=1}^{N} \begin{cases}
\big{|}  (\mathbf{D_{h}} \ubold)_{i} \big{|}_0 +  \big{|}  (\mathbf{D_{v}}\ubold)_{i}  \big{|}_0 & \text{ for } p=1,\\
\big{|} \lVert (\mathbf{D_{h}}\ubold)_{i}, (\mathbf{D_{v}}\ubold)_{i} \rVert \big{|}_0  &  \text{ for } p=2, \notag
\end{cases}
\end{align}

where by $|\cdot|_0$ we denote the function:
\begin{align*}
| z | _0:= \begin{cases}
 0 & z = 0  \\
1 & z \neq 0.
 \end{cases}
\end{align*}

The functional \eqref{eq:zero_norm} counts the number of \textit{jumps} of $\ubold$ in terms of the non-zero values of its gradient magnitude. In particular, in the case $p=1$ the regulariser independently counts the jumps along the two  horizontal and vertical Cartesian directions, whereas for $p=2$ the gradient magnitudes are taken into account jointly. In both cases, the term $\lVert \Dbold \ubold \rVert_{0,p}$ penalizes low-amplitude structures while preserving edges in the images, thus favouring sharp piece-wise constant reconstructions which are particularly desirable for image segmentation problems. We notice that $0 \leq \lVert \Dbold \ubold \rVert_{0,p} \leq 2N$ for $p\in\left\{1,2\right\}$.

In the following, we will refer to \eqref{eq:zero_norm} with $p=1$ as the \emph{anisotropic} $\ell^0$-gradient regularisation  (A-TV$^{0}$), while for $p=2$ we will refer to   \emph{isotropic} $\ell^0$-gradient regularisation (I-TV$^{0}$). 

\section{An efficient ADMM splitting} \label{sec:ADMM}
 
For $p\in\left\{1,2\right\}$, we consider the non-smooth and non-convex SR model \eqref{eq:reg_mod} with the choice \eqref{eq:zero_norm}, that is:
\begin{equation}\label{eq:min_TV0_unconstrained}
\mathbf{u^*} \in \argmin_{\ubold \in \mathbb{R}^{N}}~ \left\{\Phi(\ubold;\mu,p):=\frac{1}{2}\rVert \Sbold \Hbold \ubold - \gbold \lVert_{2}^{2} + \mu \lVert \Dbold \ubold \rVert_{0,p} \right\}.
\end{equation}
Existence of solutions for \eqref{eq:min_TV0_unconstrained} is guaranteed by the following theorem whose proof can be found in \cite[Theorem 1]{Storath14} for a general forward operator $\mathbf{A}$.

\smallskip

\begin{theorem}  \label{th:existence}
The solution set of both the anisotropic ($p=1$) and  isotropic ($p=2$)  problem \eqref{eq:min_TV0_unconstrained} is non-empty.
\end{theorem}

To solve numerically  problem \eqref{eq:min_TV0_unconstrained} we propose an iterative \emph{alternating direction method of multipliers} (ADMM) based on a suitable variable splitting. 
We separate the description for the anisotropic and isotropic case. For both cases, fixed-point convergence of the ADMM iterates upon suitable conditions is proved.

\subsection{ADMM for the anisotropic regularisation}

For $p=1$, we can rewrite the unconstrained minimisation problem \eqref{eq:min_TV0_unconstrained} in the following equivalent constrained form:
\begin{align*}
   \argmin_{\ubold}  & ~\frac{1}{2}\rVert \Sbold \Hbold \ubold - \gbold \lVert_{2}^{2}~ + ~\mu (\lVert \tbold \rVert_{0}  + \lVert \sbold \rVert_{0} ) \\  \nonumber
    s.t. \quad           & ~ \tbold:=\Dbold_{h}\ubold,\quad \sbold:=\Dbold_{v}\ubold
\end{align*}
where $\tbold,\sbold \in \mathbb{R}^{N}$ represent the horizontal/vertical gradient components, respectively.

We can then define the augmented Lagrangian function:
\begin{align} \label{eq:lagrangian_anis}
& L_{\beta_t,\beta_s}(\ubold;\tbold,\sbold,\boldsymbol{\lambda}_{t},\boldsymbol{\lambda}_{s}):=   \frac{1}{2}\rVert \Sbold \Hbold \ubold - \gbold \lVert_{2}^{2}  + \mu \lVert \tbold \rVert_{0} + \mu\lVert \sbold \rVert_{0}\nonumber \\
                          & + \langle \boldsymbol{\lambda}_{t}, \Dbold_{h} \ubold - \tbold \rangle + \dfrac{\beta_{t}}{2}\lVert \Dbold_{h}\ubold - \tbold \rVert_{2}^{2} + \langle \boldsymbol{\lambda}_{s}, \Dbold_{v}\ubold - \sbold \rangle  \nonumber\\
                          & + \dfrac{\beta_{s}}{2}\lVert \Dbold_{v}\ubold - \sbold \rVert_{2}^{2}
\end{align}
where $\beta_{t}$ and $\beta_{s}$ are two positive penalty parameters and $\boldsymbol{\lambda}_{t}$ and $\boldsymbol{\lambda}_{s}$ are the vectors of Lagrange multipliers related to the auxiliary variables $\tbold$ and $\sbold$, respectively. By letting the two parameters $\beta_t, \beta_s$ increase along the iterations (we will provide specific growth conditions in the following Theorem \ref{th:conv_ADMManiso}), we can then minimise \eqref{eq:lagrangian_anis} with respect to $\tbold,\sbold$ and $\ubold$ by iterating the following scheme:
 \begin{numcases}{}
  \scalebox{0.9}{$\tbold^{k+1}  \in  \underset{\tbold}{\argmin}~ \mu \lVert \tbold \rVert_{0} + \dfrac{\beta^{k}_{t}}{2} \lVert \tbold - (\Dbold_{h}\ubold^{k}+\dfrac{\boldsymbol{\lambda}_{t}^{k}}{\beta^{k}_{t}})\rVert_{2}^{2}$} &  \label{eq:ADMM_aniso_1} \\
  \scalebox{0.9}{$\sbold^{k+1}  \in  \underset{\sbold}{\argmin}~ \mu \lVert \sbold \rVert_{0} + \dfrac{\beta^{k}_{s}}{2} \lVert \sbold - (\Dbold_{v}\ubold^{k}+\dfrac{\boldsymbol{\lambda}_{s}^{k}}{\beta^{k}_{s}})\rVert_{2}^{2}$} & \label{eq:ADMM_aniso_2} \\
  \scalebox{0.9}{$\ubold^{k+1}  = \underset{\ubold}{\argmin} ~\dfrac{1}{2}\lVert \Sbold \Hbold \ubold - \gbold \rVert_{2}^{2} +$ } \nonumber \\
  \hspace{0cm} \scalebox{0.85}{$+ \dfrac{\beta^{k}_{t}}{2}\lVert \Dbold_{h}\ubold - (\tbold^{k+1} - \dfrac{\boldsymbol{\lambda}_{t}^{k}}{\beta^{k}_{t}})\rVert^{2}_{2} + \dfrac{\beta^{k}_{s}}{2}\lVert \Dbold_{v}\ubold - (\sbold^{k+1} - \dfrac{\boldsymbol{\lambda}_{s}^{k}}{\beta^{k}_{s}})\rVert^{2}_{2}$} & \label{eq:ADMM_aniso_3}\\
  \scalebox{0.9}{$\boldsymbol{\lambda}_{t}^{k+1} = \boldsymbol{\lambda}_{t}^{k} - \beta^{k}_{t}(\tbold^{k+1}-\Dbold_{h}\ubold^{k+1})$} &\label{eq:ADMM_aniso_4}\\
  \scalebox{0.9}{$\boldsymbol{\lambda}_{s}^{k+1} = \boldsymbol{\lambda}_{s}^{k} - \beta^{k}_{s}(\sbold^{k+1}-\Dbold_{v}\ubold^{k+1})$,} & \label{eq:ADMM_aniso_5}
  \end{numcases}
where a gradient ascent update of $\boldsymbol{\lambda}_{t}$ and $\boldsymbol{\lambda}_{s}$ is also applied. 

Under suitable growth assumptions, the sequences \eqref{eq:ADMM_aniso_1}, \eqref{eq:ADMM_aniso_2}, \eqref{eq:ADMM_aniso_3} converge to a fixed point (see Appendix \ref{sec:convergence} for the proof). 

\begin{theorem}   \label{th:conv_ADMManiso}
Let the ADMM iterations \eqref{eq:ADMM_aniso_1}-\eqref{eq:ADMM_aniso_5} be defined under the following conditions:
\begin{enumerate}[label=\textbf{A.\arabic*}]
    \item $(\beta_t^k)$,$(\beta_s^k)$ are increasing sequences such that $\sum_{k=1}^{+\infty}\sqrt{\frac{k}{\beta_t^{k}}} < + \infty$, $\sum_{k=1}^{+\infty} \sqrt{\frac{k}{\beta_s^{k}}} < + \infty$ and $  \frac{\beta_s^{k}}{\beta_t^{k}} \to c\neq 0$. \label{cond:1_ADMManiso}
    \item $\Dbold_{h}$ and $\Dbold_{v}$ are full rank.\label{cond:2_ADMManiso}
\end{enumerate}
Then, the sequences $(\tbold^{k}), (\sbold^{k}), (\ubold^{k})$ converge, i.e.:
$$
\tbold^{k} \longrightarrow \tbold^{*},\ \ 
\sbold^{k} \longrightarrow \sbold^{*},\ \
\ubold^{k} \longrightarrow \ubold^{*},
$$
with $\tbold^{*}=\Dbold_{h}\ubold^{*}$ and $\sbold^{*}=\Dbold_{v}\ubold^{*}$.
\end{theorem}
\medskip

We remark that the full rank assumption on the operators $\Dbold_h$ and $\Dbold_v$ is verified, for instance, if Dirichlet boundary conditions are assumed. A sufficient condition which guarantees the required growth of the penalty sequences is $\beta_t^k = \beta_s^k= O(k(1+\epsilon)^{k}),~ 0<\epsilon\ll 1$. 

\subsection{ADMM for the isotropic regularisation}

For $p=2$ we can write problem \eqref{eq:min_TV0_unconstrained} in the following equivalent constrained form:
\begin{align}\label{eq:iso_constrained}
    \argmin_{u} &~ \frac{1}{2}\rVert \Sbold \Hbold \ubold - \gbold \lVert_{2}^{2}~ +~ \mu  \sum_{i=1}^{N} \big{|} \lVert \zbold_{i} \rVert \big{|}_{0} \\ \nonumber
    s.t. \quad           & ~ \zbold := \Dbold \ubold
\end{align}
where $\zbold_{i}:= \big{(}(\Dbold_{h}\ubold)_{i},(\Dbold_{v}\ubold)_{i}\big{)}\in \RR^2$, for each $i=1,\ldots,N$. The augmented Lagrangian function reads in this case:
\begin{align} \label{eq:lagrangian_iso}
L_{\beta}(\ubold;\zbold,\boldsymbol{\lambda}):=  & \frac{1}{2}\rVert \Sbold \Hbold \ubold - \gbold \lVert_{2}^{2}  ~+~\mu  \sum_{i=1}^{N} \big{|} \lVert \zbold_{i} \rVert \big{|}_{0} \nonumber \\
                          & + \langle \boldsymbol{\lambda}, \Dbold \ubold - \zbold \rangle + \dfrac{\beta}{2}\lVert \Dbold \ubold - \zbold \rVert_{2}^{2}    
\end{align}
where $\beta>0$ is a scalar penalty parameter and $\boldsymbol{\lambda}\in \RR^{2\times N}$  is the Lagrange multiplier vector. As above, by letting the penalty parameter increases along the iterations at a certain growth (see the following Theorem \ref{th:conv_ADMMiso}), we seek for minimisers  of \eqref{eq:iso_constrained} by iterating the following scheme: 
\begin{numcases}{}
 \scalebox{0.85}{$\zbold^{k+1} \in \underset{\zbold}{\argmin} \ \mu\sum_{i=1}^{N} \big{|} \lVert \zbold_{i} \rVert \big{|}_{0} + \dfrac{\beta^{k}}{2} \lVert \zbold - (\Dbold \ubold^{k}+\dfrac{\boldsymbol{\lambda}^{k}}{\beta^{k}})\rVert_{2}^{2}$ }  \label{eq:ADMM_iso_1} &\\
\scalebox{0.85}{$\ubold^{k+1} = \underset{\ubold}{\argmin} \ \dfrac{1}{2}\lVert \Sbold \Hbold \ubold - \gbold \rVert_{2}^{2} + \dfrac{\beta^{k}}{2}\lVert \Dbold \ubold - (\zbold^{k+1} - \dfrac{\boldsymbol{\lambda}^{k}}{\beta^{k}})\rVert^{2}_{2}$} &  \label{eq:ADMM_iso_2} \\
\scalebox{0.85}{$\boldsymbol{\lambda}^{k+1} = \boldsymbol{\lambda}^{k} - \beta^{k}(\zbold^{k+1}-\Dbold \ubold^{k+1})$}. &  \label{eq:ADMM_iso_3}
\end{numcases}
For this scheme, a similar result as the one in Theorem \ref{th:conv_ADMManiso} holds (see Appendix \ref{sec:convergence} for a sketch of the proof).

\begin{theorem}   \label{th:conv_ADMMiso}
Let the ADMM iterations \eqref{eq:ADMM_iso_1}-\eqref{eq:ADMM_iso_2} be defined under the following conditions:
\begin{enumerate}[label=\textbf{I.\arabic*}]
    \item $(\beta^k)$ is an increasing sequence  such that $\sum_{k}^{+\infty}\sqrt{\frac{k}{\beta^{k}}} < + \infty$\label{cond:1_ADMMiso}
    \item $\Dbold$ is full rank.\label{cond:2_ADMMiso}
\end{enumerate}
Then, $(\zbold^k)\longrightarrow \zbold^{*}$, $(\ubold^k)\longrightarrow \ubold^*$ and $\zbold^{*}=\Dbold\ubold^*$.
\end{theorem}





We remark that in order to guarantee the convergence of the sequence $(\ubold^{k})$ to  $\ubold^*$,  Theorems \ref{th:conv_ADMManiso} and \ref{th:conv_ADMMiso} require full rank on the operators $\Dbold_h$ and $\Dbold_v$. This is not very limiting since Dirichlet boundary conditions can always be imposed through an artificial image padding of the image. Our numerical experiments, however, showed numerical convergence even when periodic boundary conditions are used. A theoretical convergence proof in this case is left for future research. As far as the growth condition on the penalty parameters is concerned, we remark that in \cite{Storath14} a geometric growth was assumed. Unfortunately, this is not enough for our theoretical convergence result to hold, as oscillations may appear if this is violated. We comment more on this in Section \ref{sec:numaspects}.

\subsection{Efficient solution of the ADMM subproblems} \label{subsec:subproblems}

We report here some practical details on the the efficient solutions of the subproblems \eqref{eq:ADMM_aniso_1}-\eqref{eq:ADMM_aniso_3} and  \eqref{eq:ADMM_iso_1}-\eqref{eq:ADMM_iso_2}. 

\subsubsection*{Solution of $\ell^0$ subproblems} Due to decomposability of the $\ell^{0}$ term, solving problems \eqref{eq:ADMM_aniso_1},\eqref{eq:ADMM_aniso_2} corresponds to solve the $N$ one-dimensional $\ell^2-\ell^0$ problems
\begin{align}\label{eq:1Dstandard_l0}
     \underset{\tbold_{i} \in \mathbb{R}}{\argmin}~  \delta  \lvert \tbold_{i} \rvert_{0} + ( \tbold_{i} - \fbold_{i} )_{2}^{2}
\end{align}
where $\delta=\frac{2\mu}{\beta^{k}_{t}}$ and $\fbold_i= (\Dbold_{h}\ubold^{k}+\frac{\boldsymbol{\lambda}_{t}^{k}}{\beta^{k}_{t}})_{i}$ for \eqref{eq:ADMM_aniso_1}, while $\delta=\frac{2\mu}{\beta^{k}_{s}}$, $\fbold_i= (\Dbold_{v}\ubold^{k}+\frac{\boldsymbol{\lambda}_{s}^{k}}{\beta^{k}_{s}})_i$ for \eqref{eq:ADMM_aniso_2}.  As far as the problem \eqref{eq:ADMM_iso_1} is concerned, it similarly reduces to the solution of the $N$ two-dimensional $\ell^{0}$-regularised problems
\begin{equation}\label{eq:2Dstandard_l0}
 \underset{\zbold_{i} \in \mathbb{R}^{2}}{\argmin} \  \delta  \big{|} \lVert \zbold_{i} \rVert \big{|}_{0} + \lVert \zbold_{i} - \fbold_{i} \rVert_{2}^{2}
\end{equation}
where $\delta=\frac{2\mu}{\beta^{k}}$ and $\fbold_{i} = (\Dbold_{h} \ubold^{k}_{i}+\frac{(\boldsymbol{\lambda}^{k})_{1,i}}{\beta^{k}},\Dbold_{v} \ubold^{k}_{i}+\frac{(\boldsymbol{\lambda}^{k})_{2,i}}{\beta^{k}})$.
Solving \eqref{eq:1Dstandard_l0} and \eqref{eq:2Dstandard_l0} corresponds to compute the proximal map of $|\cdot|_0$ with parameter $\delta$ evaluated in $\fbold_i$, which is nothing but the 1D \cite{blumensath2009iterative} and 2D \cite{Xu11} hard-thresholding operators, respectively.
\subsubsection*{Solution of the quadratic subproblems} 
The first order optimality conditions of problems \eqref{eq:ADMM_aniso_3} and  \eqref{eq:ADMM_iso_2} lead to the solution of large-size linear systems, whose coefficient matrix is symmetric and positive definite. To solve them efficiently, we make use of Conjugate Gradient (CG) algorithm with a warm-start initialisation at every iteration. We remark that, due to the presence of the downsampling operator $\Sbold$, the use of more efficient solvers based, for instance, on discrete Fourier transforms are here not possible, as the product matrix $\Sbold\Hbold$ does not have a block-circulant structure. However, under suitable assumptions on the down-sampling operator $\Sbold$, the problem admits a closed form solution \cite{zhao2016fast}.





\subsection{Comparisons with previous splittings}\label{sec:comparisons}
The variable splitting and the ADMM iterations considered above are different than the ones considered in \cite{Storath14,Storath15,Storath17} where the choice $\zbold=\ubold$ in \eqref{eq:iso_constrained} is made. Our choice avoids the presence of the gradient operator in the $\ell^{0}$-based problems \eqref{eq:ADMM_aniso_1}-\eqref{eq:ADMM_aniso_2} and \eqref{eq:ADMM_iso_1}, leading to the faster computation of their solution by direct solvers without requiring the use of approximate solvers based on approximate graph-cut algorithms  \cite{Storath14}. These latter algorithms have  well-known drawbacks such as strong dependence on the initialisation and require an approximate inner solver \cite{Storath15,boykov2004}. As an alternative, in \cite{Storath15}, the isotropic substep is solved by a set of anisotropic problems along the diagonal or knight-move directions, each of which is computed by dynamic programming algorithms with computational cost $O(N^{2})$ compared to $O(N)$ in our approach. 

\section{Implementation notes}  \label{sec:modelling}


\subsubsection{Operators} For the following synthetic example, we simulate the LR data  from a ground-truth HR image by applying the forward model \eqref{eq:model} where the action of the blur matrix $\Hbold$ is computed by assuming a Gaussian PSF with zero mean and standard deviation $\sigma_H$ which will be specified later on. As $\Sbold$, we consider the discretised 2D Lanczos down-sampling operator \cite{duchon1979} inbuilt in the MATLAB function \texttt{imresize}. Finally, we consider AWGN with zero mean and standard deviation $\sigma_{\bm{\eta}}$ whose values will be made precise in the following.



\subsubsection{Comparisons}  \label{sec:competitors}
We compare our results with the ones obtained by models based on gradient-sparse regularisation such as convex isotropic TV (I-TV) \cite{MarquinaOsher2008}, non-convex capped TV (c-TV) \cite{zhang2009} and anisotropic fractional TV \cite{chen2016computing} which, for consistency, have been implemented within the same ADMM optimisation framework. We further add comparisons with the results obtained by two state-of-the-art Deep Learning-based approaches. The former is the Content Adaptive Resampler (CAR) \cite{sun2020learned} convolutional neural network, which is characterised by  a downsampler-upsampler structure. For that, we use a pre-trained model \footnote{\url{https://github.com/sunwj/CAR}} taking into account only the trained upsampler part. The latter is the Image Restoration Convolutional Neural Network (IRCNN) \cite{zhang2017}, which is a Plug and Play (PnP) method based on HQS optimisation.


\subsubsection{Initialisation, parameters and evaluation metrics}
We initialise $\ubold^{0}$ in our model as $\ubold^0 = \Sbold^T \gbold$. Given the non convexity of problem \eqref{eq:min_TV0_unconstrained}, the choice of a wise initialisation is important. We tested several ones (the aforementioned one, the zero image and the I-TV initialisation) and kept the one providing the best results. 
The variables $\tbold^0,\sbold^0,\zbold^0$ as well as  $\lambdabold^0_{t},\lambdabold^0_{s},\lambdabold^0$ in \eqref{eq:ADMM_aniso_1}-\eqref{eq:ADMM_aniso_5} and in \eqref{eq:ADMM_iso_1}-\eqref{eq:ADMM_iso_3}  were set to $\mathbf{0}$. 
To ensure convergence by Theorems \ref{th:conv_ADMManiso} and \ref{th:conv_ADMMiso}, the penalty sequences are chosen as $(\beta^k)=k(1+\epsilon)^{k}$ with $\epsilon=10^{-4}$. Note that for such small choice of $\epsilon$, $ k(1+\epsilon)^{k}\approx k$, i.e. the growth of $(\beta^k)$ is almost linear. The process is stopped when the relative change between consecutive iterates $\ubold^k$ is lower than $10^{-3}$.

For simulated data, we evaluate the quality of the SR outputs by means of Peak-Signal-to-Noise-Ratio (PSNR) and  Structure Similarity index (SSIM) as well as the Jaccard index, an evaluation metric in the range $[0,1]$ measuring the ratio between correctly detected points and false detections frequently employed in the context of microscopy imaging. We remark that choosing the right evaluation metric for SR problems is not trivial, see, e.g., \cite{Sage2019} for a review. While PSNR and SSIM are good choices to quantify reconstruction quality, the Jaccard index is more appropriate to assess correct versus false pixel localisation.


\section{Numerical Experiments}\label{sec:experiments}

We report here several experiments performed on synthetic and real data. All the experiments are executed on a PC Intel(R) Core(TM) i5-6200U CPU $@$ 2.30 GHz 2.40GHz with 8.00Gb RAM using Matlab R2018b and Python 3. 
The codes are available at \url{https://github.com/pcascarano}.    

\vspace{-0.3cm}

\subsection{Computational analysis on synthetic data}\label{sec:numaspects}

We first analyse the reconstruction and the convergence properties of the proposed models/algorithms and
comment on their parameter sensitivity.

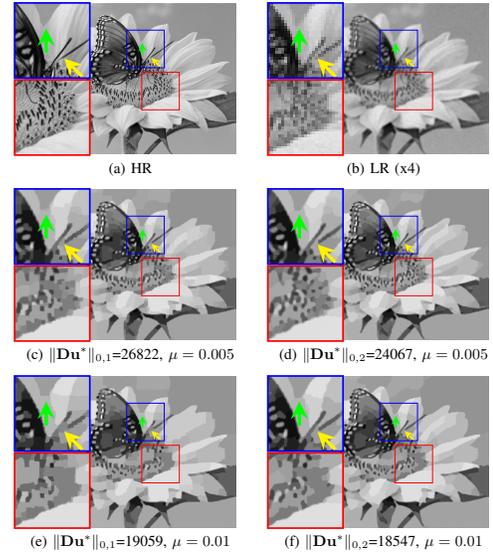
\begin{figure}	
	\centering
	\scalebox{0.8}{
	\begin{tikzpicture}
	\begin{scope}[spy using outlines={rectangle,magnification=2,size=1.25cm}]
	\node [name=c] {\includegraphics[height=2.5
	cm]{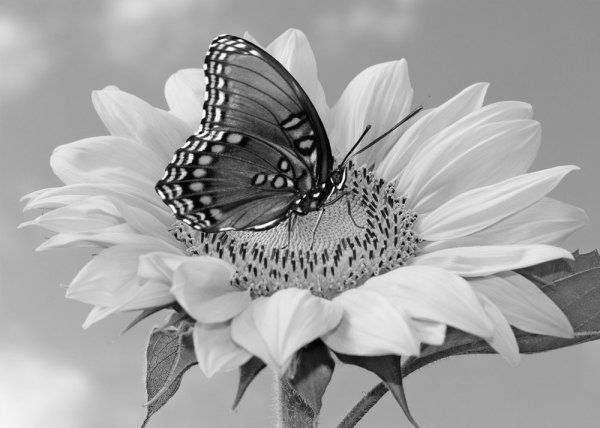}};
	\spy[red] on (0.5,-0.2) in node [name=c01] at (-1.3,-0.64);
	\spy[blue] on (0.25,0.5) in node [name=c01] at (-1.3,0.64);
	\draw [-stealth, line width=0.8pt, yellow] (0.5,0.2) -- ++(-0.15,0.15);
	\draw [-stealth, line width=0.8pt, green] (0.2,0.4) -- ++(0,0.2);
	\end{scope}
	\node[below of=c, align=center, node distance=1.5cm] {\scalebox{0.8}{\small{(a) HR}}};
	\end{tikzpicture}
	}
	\scalebox{0.8}{
	\begin{tikzpicture}
	\begin{scope}[spy using outlines={rectangle,magnification=2,size=1.25cm}]
	\node [name=c] {	\includegraphics[height=2.5cm]{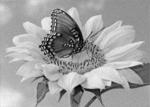}};
	\spy[red] on (0.5,-0.2) in node [name=c01] at (-1.3,-0.64);
	\spy[blue] on (0.25,0.5) in node [name=c01] at (-1.3,0.64);
	\draw [-stealth, line width=0.8pt, yellow] (0.5,0.2) -- ++(-0.15,0.15);
	\draw [-stealth, line width=0.8pt, green] (0.2,0.4) -- ++(0,0.2);
	\end{scope}
	\node[below of=c, align=center, node distance=1.5cm] {\scalebox{0.8}{\small{(b) LR (x4)}}};
	\end{tikzpicture}
		}
	\scalebox{0.8}{
    \begin{tikzpicture}
    \begin{scope}[spy using outlines={rectangle,black,magnification=2,size=1.25cm}]
	\node [name=c1] {	\includegraphics[height=2.5cm]{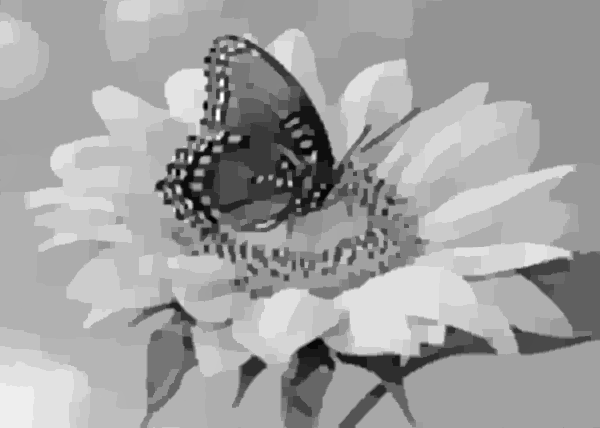}};
	\spy[red] on (0.5,-0.2) in node [name=c11] at (-1.3,-0.64);
	\spy[blue] on (0.25,0.5) in node [name=c11] at (-1.3,0.64);
	\draw [-stealth, line width=0.8pt, yellow] (0.5,0.2) -- ++(-0.15,0.15);
	\draw [-stealth, line width=0.8pt, green] (0.2,0.4) -- ++(0,0.2);
	\end{scope}
	\node[below of=c1, align=center, node distance=1.5cm] {\scalebox{0.8}{\small{(c) $\lVert\Dbold\ubold^*\rVert_{0,1}$}=26822, $\mu=0.005$}};
	\end{tikzpicture}
	}
	\scalebox{0.8}{
    \begin{tikzpicture}
	\begin{scope}[spy using outlines={rectangle,black,magnification=2,size=1.25cm}]
	\node [name=c2] {	\includegraphics[height=2.5cm]{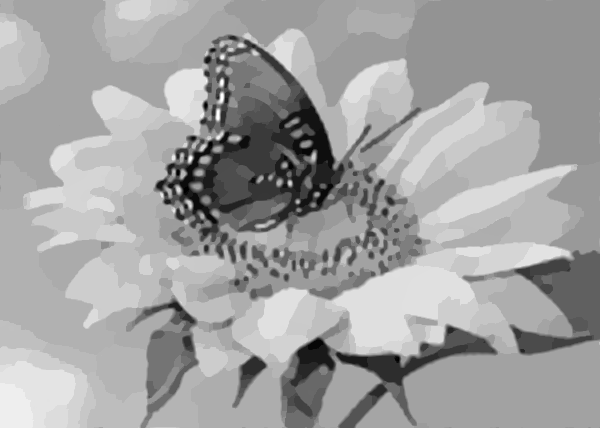}};
	\spy[red] on (0.5,-0.2) in node [name=c21] at (-1.3,-0.64);
	\spy[blue] on (0.25,0.5) in node [name=c21] at (-1.3,0.64);
	\draw [-stealth, line width=0.8pt, yellow] (0.5,0.2) -- ++(-0.15,0.15);
	\draw [-stealth, line width=0.8pt, green] (0.2,0.4) -- ++(0,0.2);
	\end{scope}
	\node[below of=c2, align=center, node distance=1.5cm] {\scalebox{0.8}{\small{(d) $\lVert\Dbold\ubold^*\rVert_{0,2}$}=24067, $\mu=0.005$}};
	\end{tikzpicture}
	}
	\scalebox{0.8}{
	\begin{tikzpicture}
	\begin{scope}[spy using outlines={rectangle,black,magnification=2,size=1.25cm}]
	\node[name=c3]{	\includegraphics[height=2.5cm]{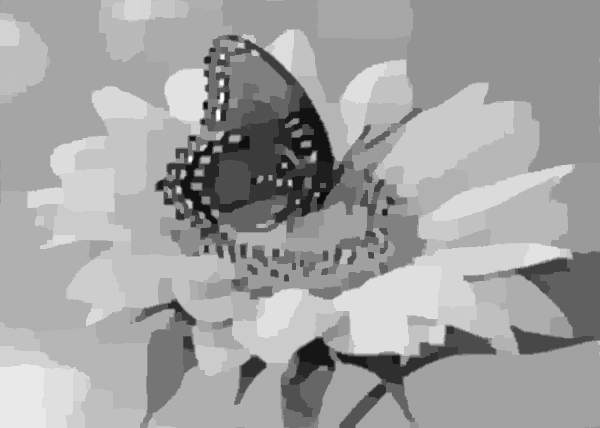}};
	\spy[red] on (0.5,-0.2) in node [name=c31] at (-1.3,-0.64);
	\spy[blue] on (0.25,0.5) in node [name=c31] at (-1.3,0.64);
	\draw [-stealth, line width=0.8pt, yellow] (0.5,0.2) -- ++(-0.15,0.15);
	\draw [-stealth, line width=0.8pt, green] (0.2,0.4) -- ++(0,0.2);
	\end{scope}
	\node[below of=c3, align=center, node distance=1.5cm] {\scalebox{0.8}{\small{(e) $\lVert\Dbold\ubold^*\rVert_{0,1}$}=19059, $\mu = 0.01$}};
	\end{tikzpicture}
	}
	\scalebox{0.8}{
	\begin{tikzpicture}
	\begin{scope}[spy using outlines={rectangle,black,magnification=2,size=1.25cm}]
	\node[name=c4] {	\includegraphics[height=2.5cm]{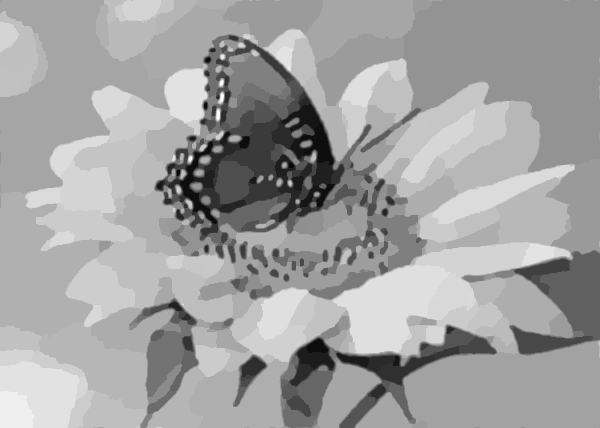}};
	\spy[red] on (0.5,-0.2) in node [name=c41] at (-1.3,-0.64);
	\spy[blue] on (0.25,0.5) in node [name=c41] at (-1.3,0.64);
	\draw [-stealth, line width=0.8pt, yellow] (0.5,0.2) -- ++(-0.15,0.15);
	\draw [-stealth, line width=0.8pt, green] (0.2,0.4) -- ++(0,0.2);
	\end{scope}
	\node[below of=c4, align=center, node distance=1.5cm] {\scalebox{0.8}{\small{(f) $\lVert\Dbold\ubold^*\rVert_{0,2}$}=18547, $\mu = 0.01$}};
	\end{tikzpicture}
	}
	
	\caption{\small{Results obtained for  $\mu\in\left\{0.005,0.1\right\}$ by the SR anisotropic (A-TV$^0$) and isotropic (I-TV$^0$) $\ell^0$ gradient-sparse model on a synthetic image.}}
	
	\label{fig:butterfly}
    
    \end{figure}

For this first example, LR data were generated by applying  \eqref{eq:model} to the HR  $428 \times 600$ grayscale image in Figure \ref{fig:butterfly} (a). Gaussian blur with $\sigma_{H}=1$ and down-sampling with factor $L=4$ were applied and AGWN with standard deviation $\sigma_{\eta}=0.01$ was added to get the LR image in Figure \ref{fig:butterfly} (b). 
In Figure \ref{fig:butterfly} (c)-(f) we report the results computed  by the anisotropic (A-$\text{TV}^0$) and isotropic (I-$\text{TV}^0$) $\ell^0$-gradient model for two different values of the regularisation parameter $\mu\in\left\{0.005,0.01\right\}$.
The jump-sparse regularisation flattens out many details in the reconstruction, promoting a  cartoon-like reconstruction which can then be used for subsequent classification and segmentation purposes: the higher the regularisation parameter $\mu$, the more simplified the reconstruction. We further add a close-up of two ROIs: the blue square contains both fine details (filaments, yellow arrows) and corner points (green arrows), the red one textured details.
The directional bias of the A-TV$^{0}$ regularisation along the horizontal and vertical direction is here clearly visible.  
We report in the captions of the Figure \ref{fig:butterfly} (c)-(f) the values $\lVert\Dbold\ubold^*\rVert_{0,1}$ and $\lVert\Dbold\ubold^*\rVert_{0,2}$ which corresponds to the number of gradient jumps on the output image. Note that choosing a larger $\mu$, more jump-sparsity is promoted so the number of jumps on $\ubold^*$ is smaller.

\begin{figure}[ht!]
	\centering
	\subfloat[]{{\includegraphics[width=0.24\textwidth]{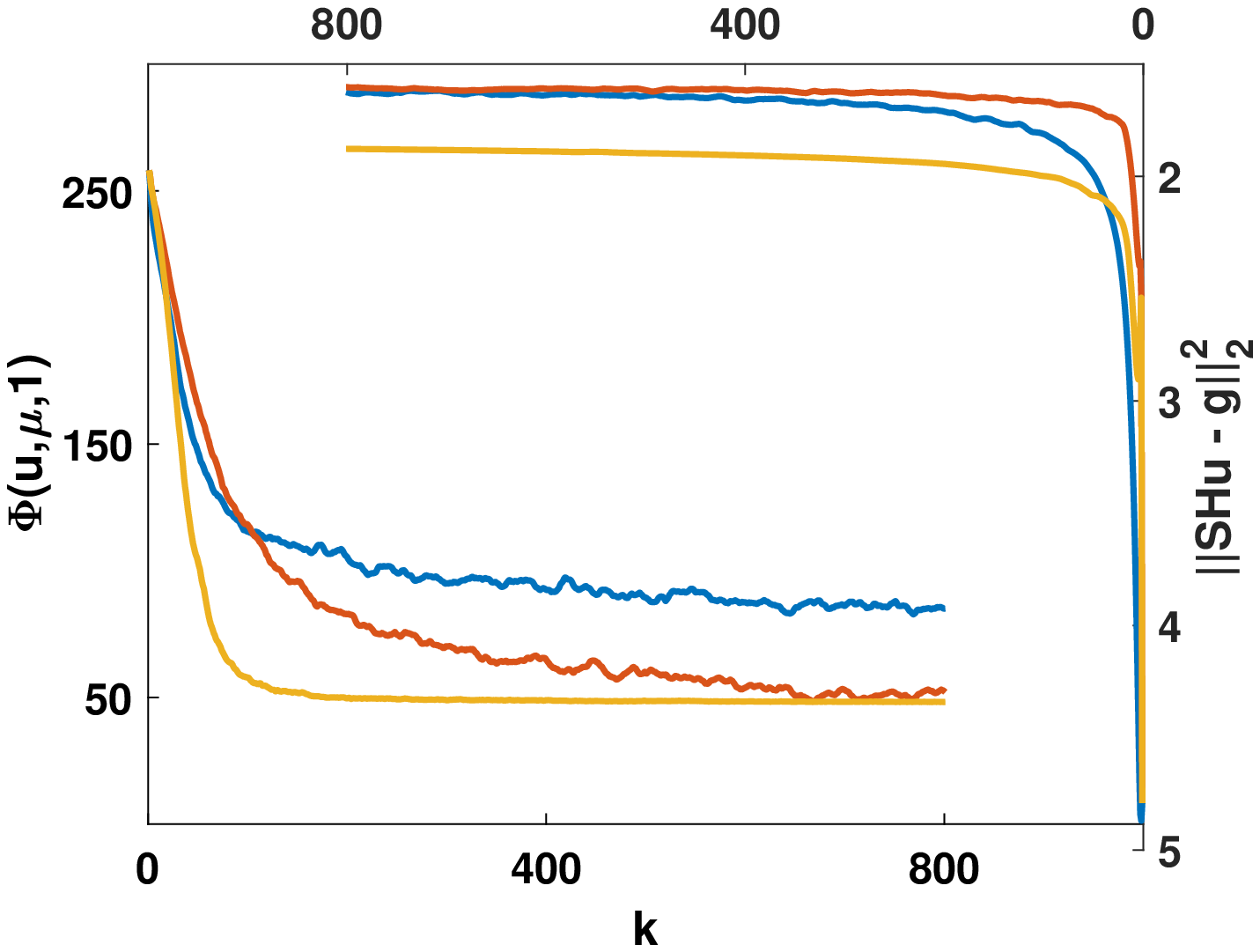}}}
	\subfloat[]{{\includegraphics[width=0.24\textwidth]{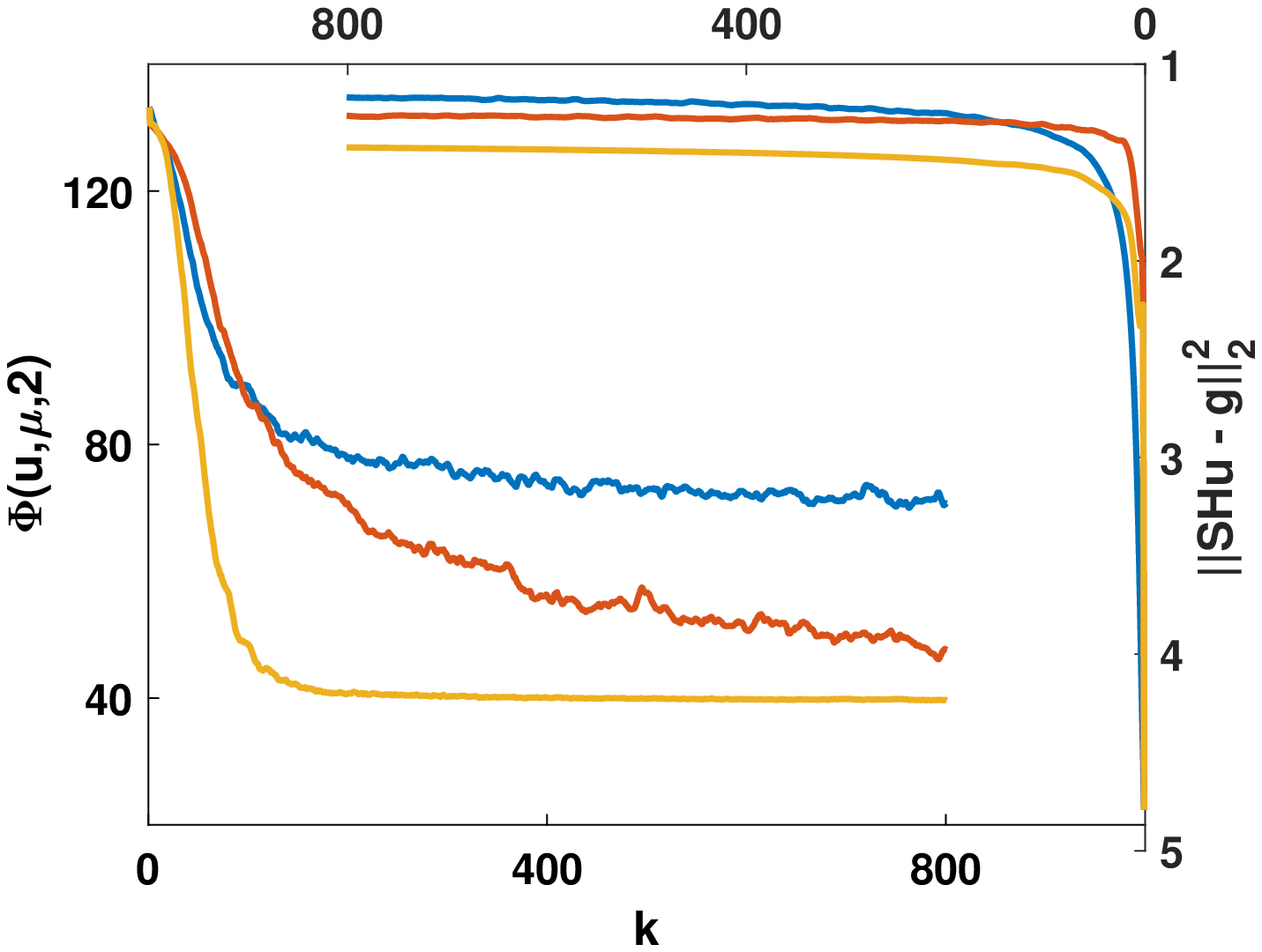}}}
	\caption{\small{Values of the cost function in \eqref{eq:min_TV0_unconstrained} (left $y$-axis) and of the fidelity term (right $y$-axis)  along iterations in the two cases $\Phi (\ubold^k;\mu,1)$ (a) and $\Phi (\ubold^k;\mu,2)$ (b). The penalty sequences are chosen as  $\beta^{k}=\beta^k_t=\beta^k_s\equiv10$ (blue), $\beta^{k}=\beta^k_t=\beta^k_s=k^{0.5}$ (red), $\beta^{k}=\beta^k_t=\beta^k_s=k(1 + \epsilon)^{k}$} with $\epsilon=10^{-4}$ (yellow).}
	\label{fig:beta1}
\end{figure}

We now validate the algorithmic convergence behaviour w.r.t. to the choice of the penalty sequences $(\beta_t^k), (\beta_s^k), (\beta_k)$.  Namely, in Figure \ref{fig:beta1} (a)  and \ref{fig:beta1} (b) we report the behaviour of the objective functions $\Phi(\ubold^k;\mu,p)$ in \eqref{eq:min_TV0_unconstrained} along the ADMM iterations for different choices of the penalty sequences (left).  For both cases $p=1$ and $p=2$ we choose $\beta^{k}=\beta^k_t=\beta^k_s\equiv10$ for all $k$ (blue line), $\beta^{k}=\beta^k_t=\beta^k_s=k^{0.5}$ (red line) and $\beta^{k}=\beta^k_t=\beta^k_s=k(1 + \epsilon)^{k}$ with $\epsilon=10^{-4}$ (yellow line). On the same plots we further show the decay of the quadratic data term (right). We observe that  when the penalty sequence fulfil the required growth condition then the convergence is nicely monotone. whereas for the other two choices, the decay exhibits oscillations while preserving a globally decreasing trend. Numerically, this suggests that possibly less severe growth conditions may be employed, such as a sufficiently large constant values of the penalty parameters. A further study on this is left for future research. 


To confirm the improved computational performance of our ADMM algorithm w.r.t.~  to the one proposed in \cite{Storath15} and adapted to solve the SR problem \eqref{eq:min_TV0_unconstrained}, we report in Table \ref{tab:4} a comparison table both in terms of number of iterations-to-convergence and  computational times. We stress that the poor performance of the ADMM algorithm in \cite{Storath15} is due here to the large computational cost required to solve the $\ell^0$ gradient steps via inner optimisation routines.  This, combined with the use of CG solvers (required for the SR problem under consideration as no Fourier-based approaches can be used in general) makes the overall cost much higher in comparison to our more explicit splitting.

\begin{table}[t] 
\begin{center}
\caption{\small{Iterations till convergence (\texttt{iter}) and computational time (in seconds) for different methods solving \eqref{eq:min_TV0_unconstrained}.}} 
\scalebox{0.9}{
\begin{tabular}{c|cccc}\hline
\textbf{Method}  & \cite{Storath15} & A-TV$^{0}$ & I-TV$^{0}$  \\ \hline 
\texttt{iter}   & 1905   & 63 & 59 \\
time (s) & 2866.31   & 214.83  & 195.99 \\ 
\hline
\end{tabular} 
}
\label{tab:4}
\end{center}
\end{table}

\vspace{-0.5cm}

\subsection{Real-world applications} \label{sec:Applications}
We now report the results obtained by applying the proposed model to different real-world applications where a SR version of the given LR image is required for further image analysis.

\subsubsection{QR code recognition} \label{sec:QR}

The first application we consider is the problem of QR super-resolution. As described in, e.g., \cite{Kato2011}, images of QR codes are often scanned nowadays by means of portable devices with limited resolution. Furthermore, QR scans are often taken from a distance and in non-optimal optical conditions so that blur and noise  further limit the amount of visible information, thus making SR desirable. 

For our tests, we first generate a binary QR code image of size $250 \times 250$ by using a free QR code generator \footnote{\url{https://www.qrme.co.uk/}}, then we simulate several LR acquisitions for different levels of degradation. We consider three test cases: $\sigma_{\eta}=0.01$ and $\sigma_H=1$ (TEST 1), $\sigma_{\eta}=0.05$ and $\sigma_H=1$ (TEST 2) and $\sigma_{\eta}=0.01$ and $\sigma_H=4$ (TEST 3). We compare the results obtained by our model with the ones obtained by the models in Section \ref{sec:competitors}. For each method, we select the model parameters maximising the Jaccard index. 
Furthermore, to avoid non-binary outputs (required for Jaccard index computations), we post-process the SR results by means of an adaptive Otsu thresholding and re-compute the evaluation metrics on the binarised output, see Table \ref{tab:2}.

\begin{figure}[ht!] 
	\centering
    
    \begin{tikzpicture}

    \begin{scope}[spy using outlines={rectangle,red,magnification=2,width=1cm, height=2cm}]
	\node {\includegraphics[height=1cm]{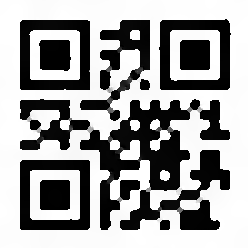}};
	\spy on (-0.25,0) in node [name=TV0I]  at (4.14,-2.4);
	\end{scope}
	\node[below of=TV0I, align=center, node distance=1.2cm] {\scalebox{0.70}{I-TV$^{0}$}};

	\begin{scope}[spy using outlines={rectangle,blue,magnification=2,width=1cm, height=2cm}]
	\node {	\includegraphics[height=1cm]{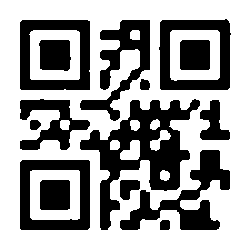}};
	\spy on (0.25,0) in node [name=TV0I_Bin]  at (5.17,-2.4);
	\end{scope}
	\node[below of=TV0I_Bin, align=center, node distance=1.2cm] {\scalebox{0.70}{I-TV$^{0}$-B}};
	
	\begin{scope}[spy using outlines={rectangle,blue,magnification=2,width=1cm, height=2cm}]
	\node {	\includegraphics[height=1cm]{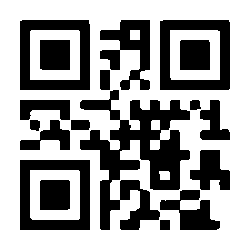}};
	\spy on (0.25,0) in node [name=TV0A_Bin]  at (3.10,-2.4);
	\end{scope}
	\node[below of=TV0A_Bin, align=center, node distance=1.2cm] {\scalebox{0.70}{A-TV$^{0}$-B}};
	
	\begin{scope}[spy using outlines={rectangle,red,magnification=2,width=1cm, height=2cm}]
	\node {\includegraphics[height=1cm]{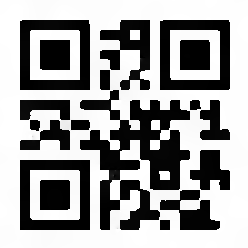}};
	\spy on (-0.25,0) in node [name=TV0A]  at (2.07,-2.4);
	\end{scope}
	\node[below of=TV0A, align=center, node distance=1.2cm] {\scalebox{0.70}{A-TV$^{0}$}};

    \begin{scope}[spy using outlines={rectangle,blue,magnification=2,width=1cm, height=2cm}]
	\node {	\includegraphics[height=1cm]{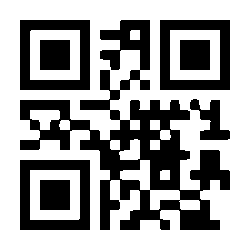}};
	\spy on (0.25,0) in node [name=Zhang_Bin]  at (1.03,-2.4);
	\end{scope}
	\node[below of=Zhang_Bin, align=center, node distance=1.2cm] {\scalebox{0.70}{IRCNN-B}};
	
    \begin{scope}[spy using outlines={rectangle,red,magnification=2,width=1cm, height=2cm}]
	\node {\includegraphics[height=1cm]{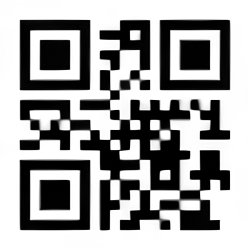}};
	\spy on (-0.25,0) in node [name=Zhang]  at (0,-2.4);
	\end{scope}
	\node[below of=Zhang, align=center, node distance=1.2cm] {\scalebox{0.70}{IRCNN}};

	\begin{scope}[spy using outlines={rectangle,blue,magnification=2,width=1cm, height=2cm}]
	\node {	\includegraphics[height=1cm]{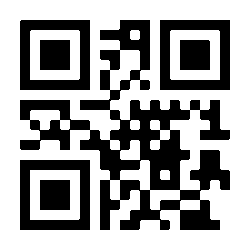}};
	\spy on (0.25,0) in node [name=EDSR_Bin]  at (5.17,0);
	\end{scope}
	\node[below of=EDSR_Bin, align=center, node distance=1.2cm] {\scalebox{0.70}{CAR-B}};
	
	\begin{scope}[spy using outlines={rectangle,red,magnification=2,width=1cm, height=2cm}]
	\node {\includegraphics[height=1cm]{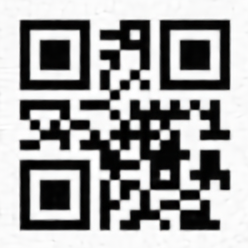}};
	\spy on (-0.25,0) in node [name=EDSR]  at (4.14,0);
	\end{scope}
	\node[below of=EDSR, align=center, node distance=1.2cm] {\scalebox{0.70}{CAR}};

	\begin{scope}[spy using outlines={rectangle,blue,magnification=2,width=1cm, height=2cm}]
	\node {	\includegraphics[height=1cm]{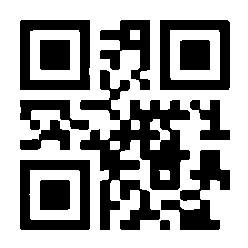}};
	\spy on (0.25,0) in node [name=cL1_Bin]  at (3.10,0);
	\end{scope}
	\node[below of=cL1_Bin, align=center, node distance=1.2cm] {\scalebox{0.70}{A-TV$^{1/2}$-B}};
	
	\begin{scope}[spy using outlines={rectangle,red,magnification=2,width=1cm, height=2cm}]
	\node {\includegraphics[height=1cm]{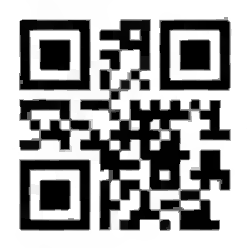}};
	\spy on (-0.25,0) in node [name=cL1]  at (2.07,0);
	\end{scope}
	\node[below of=cL1, align=center, node distance=1.2cm] {\scalebox{0.70}{A-TV$^{1/2}$}};

	\begin{scope}[spy using outlines={rectangle,blue,magnification=2,width=1cm, height=2cm}]
	\node {	\includegraphics[height=1cm]{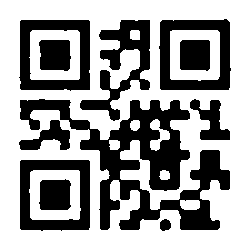}};
	\spy on (0.25,0) in node [name=TV_Bin]  at (1.03,0);
	\end{scope}
	\node[below of=TV_Bin, align=center, node distance=1.2cm] {\scalebox{0.70}{I-TV-B}};
	
	\begin{scope}[spy using outlines={rectangle,red,magnification=2,width=1cm, height=2cm}]
	\node {\includegraphics[height=1cm]{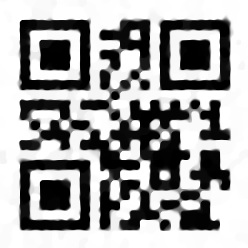}};
	\spy on (-0.25,0) in node [name=TV]  at (0,0);
	\end{scope}
	\node[below of=TV, align=center, node distance=1.2cm] {\scalebox{0.70}{I-TV}};
	
	\end{tikzpicture}
    
    \caption{\small{QR SR results obtained by different methods on TEST2 image before (red frames) and after (blue frames) binarisation.}}
	\label{fig:QRcode1}
\end{figure}
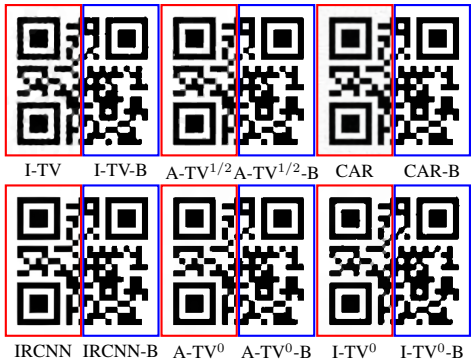

\begin{figure}[ht!]
	\centering
	\scalebox{0.8}{
	\begin{tikzpicture}

	\begin{scope}[spy using outlines={rectangle,black,magnification=2,size=2cm}]
	\node {	\includegraphics[height=2cm]{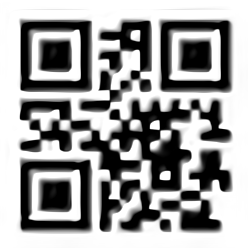}};
	\spy on (-0.4,-0.4) in node [name=c6]  at (2.03,0);
	\end{scope}
	\node[below of=c6, align=center, node distance=1.2cm] {\scalebox{0.8}{IRCNN}};
	
	\begin{scope}[spy using outlines={rectangle,black,magnification=2,size=2cm}]
	\node {	\includegraphics[height=2cm]{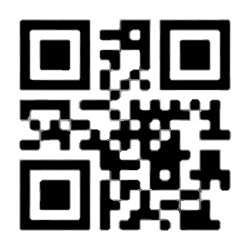}};
	\spy on (-0.4,-0.4) in node [name=c2]  at (4.03,0);
	\end{scope}
	\node[below of=c2, align=center, node distance=1.2cm] {\scalebox{0.8}{A-TV$^{0}$}};

	\begin{scope}[spy using outlines={rectangle,black,magnification=2,size=2cm}]
	\node {	\includegraphics[height=2cm]{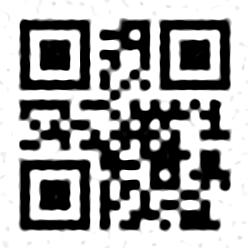}};
	\spy on (-0.4,-0.4) in node [name=c1]  at (-1.97,0);
	\end{scope}
	\node[below of=c1, align=center, node distance=1.2cm] {\scalebox{0.8}{I-TV}};
	
	\begin{scope}[spy using outlines={rectangle,black,magnification=2,size=2cm}]
	\node {	\includegraphics[height=2cm]{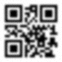}};
	\spy on (-0.4,-0.4) in node [name=c] at (-4,0);
	\end{scope}
	\node[below of=c, align=center, node distance=1.2cm] {\scalebox{0.8}{LR (x4)}};
	
	\begin{scope}[spy using outlines={rectangle,black,magnification=2,size=2cm}]
	\node {	\includegraphics[height=2cm]{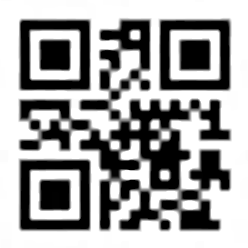}};
	\spy on (-0.4,-0.4) in node [name=c1]  at (0.03,0);
	\end{scope}
	\node[below of=c1, align=center, node distance=1.2cm] {\scalebox{0.8}{c-TV}};
		
	
	\end{tikzpicture}
	}
	\caption{\small{Details of QR SR outputs obtained by different methods on TEST 3 image.} }
	\label{fig:QRcode2}
	
\end{figure}
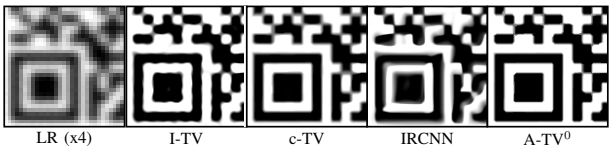

In Figure \ref{fig:QRcode1} we report the results obtained by the different methods for the TEST 2 image before (red frame) and after (blue frame) binarisation. We observe that due to the sharp nature of the the TV$^{0}$ regularisations, the results are almost binary so they do not benefit much from the post-processing step in terms of Jaccard index values as the other methods do. In Figure \ref{fig:QRcode2} we report a zoom of the best results obtained before binarisation by all  methods starting from the TEST 3 highly corrupted LR image. 

\vspace{-0.5cm}

The quantitative evaluation of the results in terms of PSNR, SSIM and Jaccard index for the three different test cases is reported in Table \ref{tab:2}. Without any binarisation, the A-TV$^{0}$ model outperforms all the others as far as the PSNR, SSIM and Jaccard indices are concerned. The simplified geometry of the QR images considered (i.e. the sole presence of horizontal/vertical edges) makes in fact this kind of data tailored for such geometrically-biased regularisations. Furthermore, the highly non-convex jump-sparsification forces the ouptut to be almost  binary, without the need of any further post-processing binarisation, as it is required by all the other regularisations to achieve comparable (if not better) quality scores. This simple example shows that the image simplification intrinsically favoured by the use of TV$^{0}$ regularisers shall limits the need of post-processing techniques. 

As far as the deep-learning results are concerned,
we remark that the CAR network in this experiment is used in a transfer learning mode, with no noisy nor blurred images observed in the training phase.
For a fairer comparison, we thus consider the IRCNN PnP network which is capable to handle different levels of degradations, although it is shown to fail in the presence of highly-degraded data, see Figure \ref{fig:QRcode2}.

\begin{table}[h] 
\begin{center}
\caption{\small{Quantitative evaluation of SR models performance on QR  for three different TEST images and methods.
By ``-B" we denote results after binarisation. In each column we colour red the best method, blue the second-best.}} 
\scalebox{0.9}{
\begin{tabular}{c|cccccc}\hline
\textbf{LR}                        & \textbf{Method} & \textbf{PSNR} & \textbf{PSNR-B} & \textbf{SSIM}  & \textbf{SSIM-B} & \textbf{Jaccard} \\ \hline 
\multirow{6}{*}{\textbf{TEST 1}}                     & I-TV$^{0}$   & 22.5199   & 29.0809   & 0.9423   & 0.9873  & 0.9980   \\ 
                           & A-TV$^{0}$ & \textcolor{red}{32.5943}   & \textcolor{blue}{35.8478}   & \textcolor{red}{0.9913}   & \textcolor{blue}{0.9989}  & \textcolor{red}{0.9999}   \\ 
                          & I-TV    & 23.3845   & 26.3357   & 0.9489   & 0.9762  & 0.9963   \\ 
                          & c-TV    & 19.4522   & \textcolor{red}{36.7496}   & 0.8849   & 0.9977  & \textcolor{blue}{0.9997}  \\
                         & A-TV$^{1/2}$    & 18.6328   & \textcolor{red}{36.7496}   & 0.8594   & 0.9989  & \textcolor{blue}{0.9997}  \\
                           & CAR   & 20.2460   & 27.8163   &  0.8159  & 0.9801  & 0.9966    \\
                          & IRCNN  & \textcolor{blue}{25.0589}   & 35.3363   & \textcolor{blue}{0.9622}   & \textcolor{red}{0.9992} & 0.9995   \\ \hline
\multirow{6}{*}{\textbf{TEST 2}}                           & I-TV$^{0}$   & 19.3318   & 18.6308   & 0.8766   & 0.9156  & 0.9781   \\ 
                          & A-TV$^{0}$   & \textcolor{red}{22.6887}   & 22.6256   & \textcolor{red}{0.9242}   & 0.9653  & \textcolor{red}{0.9912}   \\ 
                          & I-TV    & 18.1101   & 18.9848   & 0.8012   & 0.9171  & 0.9798   \\ 
                          & c-TV    & 18.7331   & 21.3473   & 0.8211   & 0.9595  & 0.9882   \\
                          & A-TV$^{1/2}$   & 19.2182   & 22.5108   & 0.8664   & 0.9660  & \textcolor{blue}{0.9910}  \\
                          & CAR   & 18.1320   & \textcolor{red}{26.7831}   &  0.7493  & \textcolor{blue}{0.9805}  & 0.9906    \\
                          & IRCNN  & \textcolor{blue}{21.4314}   & \textcolor{blue}{26.3968}   & \textcolor{blue}{0.9057}  & \textcolor{red}{0.9850}  & 0.9902   \\ \hline
                          
\multirow{6}{*}{\textbf{TEST 3}}     & I-TV$^{0}$  & \textcolor{blue}{18.3763}   & 19.7532   & \textcolor{blue}{0.8634}   & 0.9294  & 0.9831  \\ 
                          & A-TV$^{0}$  & \textcolor{red}{19.2908}   & \textcolor{blue}{21.9341}   & \textcolor{red}{0.8861}   & \textcolor{blue}{0.9556}  & \textcolor{blue}{0.9897}  \\ 
                          & I-TV    & 17.9552   & 20.1585   & 0.8222   & 0.9282  & 0.9846   \\ 
                          & c-TV    & 16.9580   & \textcolor{red}{22.4648}   & 0.7915   & \textcolor{red}{0.9605}  & \textcolor{red}{0.9917}   \\
                          &  A-TV$^{1/2}$    & 17.0785  & 20.6874   & 0.7706   & 0.9372  & 0.9863   \\
                          & CAR   & 11.1809   & 11.5412   &  0.4057  & 0.6342  & 0.8887    \\
                          & IRCNN  & 14.2915   & 12.5640   & 0.6342   & 0.6565  & 0.9133   \\ \hline
\end{tabular} 
}
\label{tab:2}
\end{center}
\end{table}

\subsubsection{Land-cover classification} \label{sec:landcover} 
The exploitation of Multi-Spectral Images (MSIs) is fundamental in the field of land-cover mapping and classification \cite{cihlar2000}. MSIs are satellite images whose numerous channels (from 4 to 200) are acquired at a different electromagnetic spectrum bandwidth, such as visible or infrared bands, which quantifies different types of information about the objects in the recorded scene, such as their physical composition and their temperature. Existing  segmentation techniques exploit these properties to label each pixel of the MSI within a class, thus producing a final 2D labelled image. These maps are essential in many sustainability-related applications and monitoring purposes for detecting land-cover changes (e.g. deforestation) over the years at the same geographical location, which cannot be done directly by simply looking at the MSIs (see \cite{malkin2018} and references therein). Among the many existing open-source MSI datasets, we consider here e.g. the National Agriculture Imagery Program (NAIP)  \cite{maxwell2017} dataset and the Hamlin Beach State Park (HBSP)  \cite{kemker2017} dataset. The former is a collection of HR MSIs with 1 meter resolution and three RGB channels. The latter is a database of MSIs with 6 channels, 3 for the RGB and 3 for the infrared bands and is used for validating the performance of deep-learning-based segmentation algorithms aiming to differentiate land objects with analogous characteristics (e.g. a grass from a tree), see \cite{kemker2017}. %

For this problem, we apply SR methods to increase the spatial resolution of the given MSI image so as to produce an output image which could be easily segmented by standard segmentation algorithms. The need of a SR model in this specific application is justified by the physical limitations preventing HR acquisitions, such as the limited spatial resolution in some bandwidths as the infrared band \cite{mandanici2020}. On the other hand, a simplified image where noise and blur artefacts are removed comes very handy for classification purposes.  To compute the land-cover mapping on the output of the SR regularised images we use in the following a standard $k$-Means segmentation and the state-of-the-art U-Net neural network \cite{Ronneberger2015}, specifically developed for segmentation tasks. 

%


\begin{figure}
	\centering
	\scalebox{0.7}{
	\begin{tikzpicture}
	\begin{scope}[]
	\node [name=c] {\includegraphics[height=4
	cm]{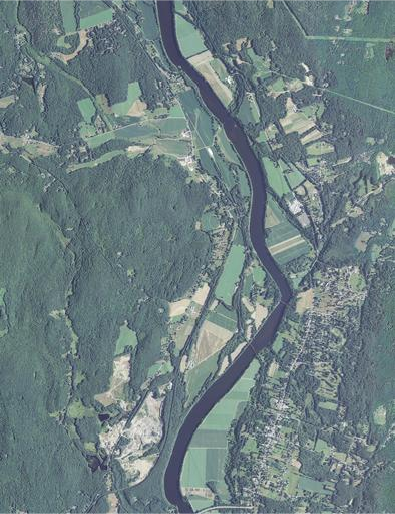}};
	\end{scope}
	\node[below of=c, align=center, node distance=2.2cm] {\small{(a) NAIP LR}};
	\end{tikzpicture}\begin{tikzpicture}
	\begin{scope}[]
	\node [name=c] {\includegraphics[height=4
	cm]{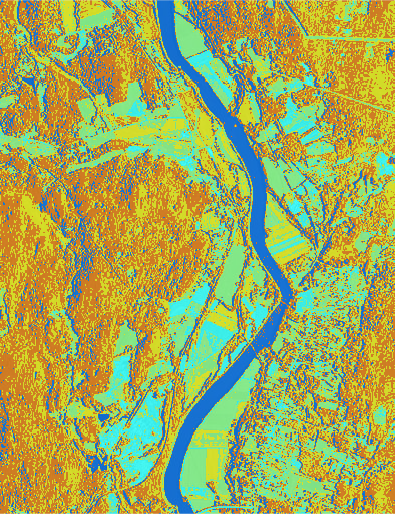}};
	\end{scope}
	\node[below of=c, align=center, node distance=2.2cm] {\small{(b) $k$-MEANS LR (x2)}};
	\end{tikzpicture}	
	}
	
     \centering
     \scalebox{0.7}{
	\begin{tikzpicture}
	\begin{scope}[]
	\node [name=c] {\includegraphics[height=4
	cm]{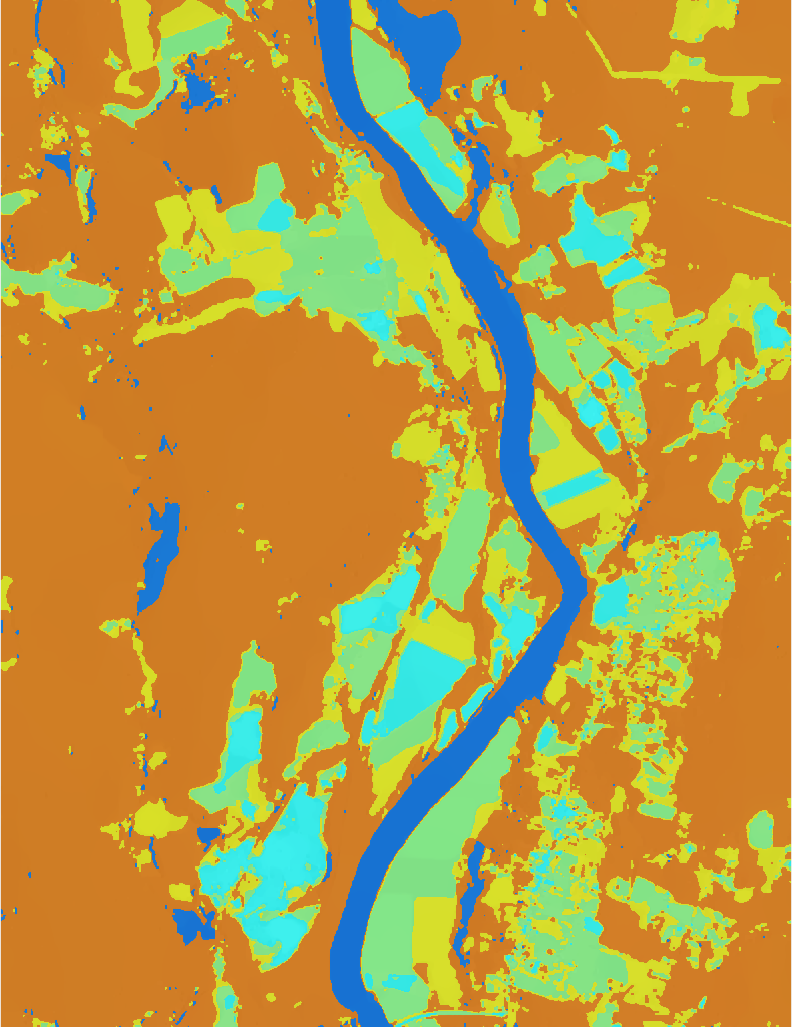}};
	\draw[red,line width=0.8mm,dotted] (-0.5,-0.8) rectangle (-1.2, 0.4);
	\draw[red,line width=0.8mm,dotted] (-0.2,1.3) rectangle (0.5, 1.95);
	\end{scope}
	\node[below of=c, align=center, node distance=2.2cm] {\small{(c) $k$-MEANS I-TV} };
	\end{tikzpicture}\begin{tikzpicture}
	\begin{scope}[]
	\node [name=c] {\includegraphics[height=4
	cm]{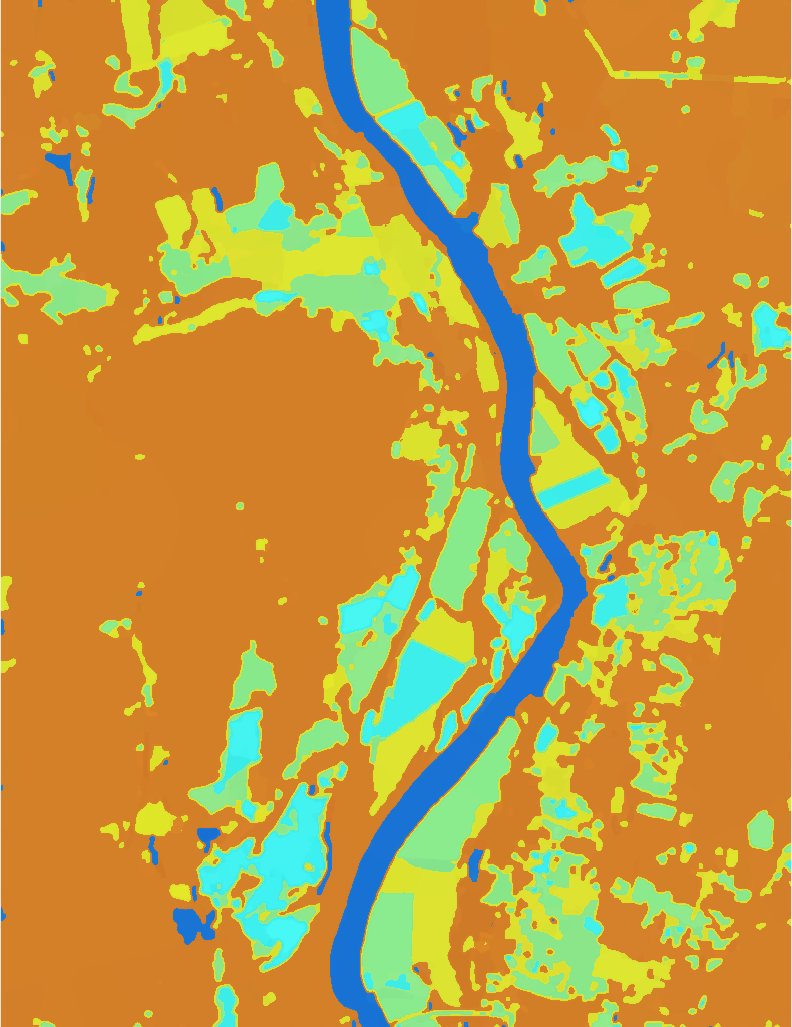}};
	\draw[green,line width=0.8mm,dotted] (-0.5,-0.8) rectangle (-1.2, 0.4);
	\draw[green,line width=0.8mm,dotted] (-0.2,1.3) rectangle (0.5, 1.95);
	\end{scope}
	\node[below of=c, align=center, node distance=2.2cm] {\small{(d) $k$-MEANS I-TV$^0$}};
	\end{tikzpicture}
	}
	\caption{\small{$k$-means segmentation ($k=5$) of MSI data. In (c)-(d) the red and green boxes show possible misclassified details. (a) LR  image (x2)  (b) $k$-means classification of LR image (c) $k$-means classification of I-TV output (d) $k$-means classification of I-TV$^0$ output.}}
	\label{fig:landcover1}

    \end{figure}
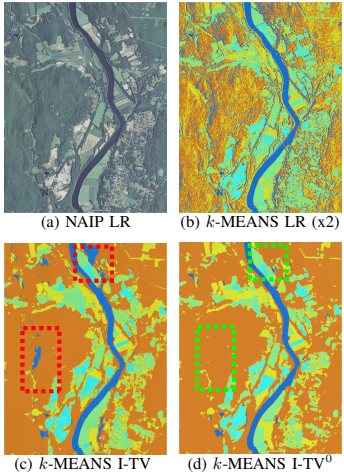







In the first experiment we consider a LR test image \footnote{Image identification number: M 4207221 NW 18 1 20120709} from the NAIP dataset (Figure \ref{fig:landcover1} (a)). We first run the $k$-Means algorithm directly on this image, choosing empirically the number of classes to be $k=5$. The classification obtained looks speckled and significant classification errors occur (see \ref{fig:landcover1} (b)). In Figures \ref{fig:landcover1} (c)-(d), we report the classification results obtained by applying  $k$-Means to the I-TV and I-TV$^{0}$ SR reconstructions (with $L=2$). 
The segmentation results obtained on these gradient-sparsified images appear much more reliable. We notice, in particular, that some parts of the vegetation are wrongly labelled as water in the I-TV result (red boxes), whereas this is not the case for the I-TV$^{0}$ reconstruction (green boxes), due to its enhanced flattening properties.  

\begin{figure}
	\centering
	\subfloat[]{{\includegraphics[width=0.14\textwidth]{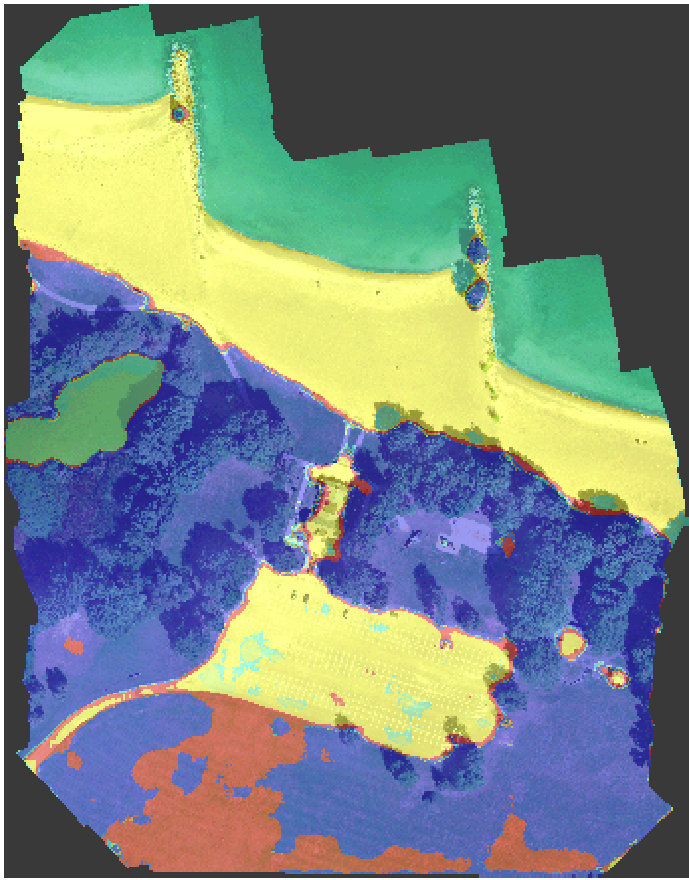}}} \quad
	\subfloat[]{{\includegraphics[width=0.14\textwidth]{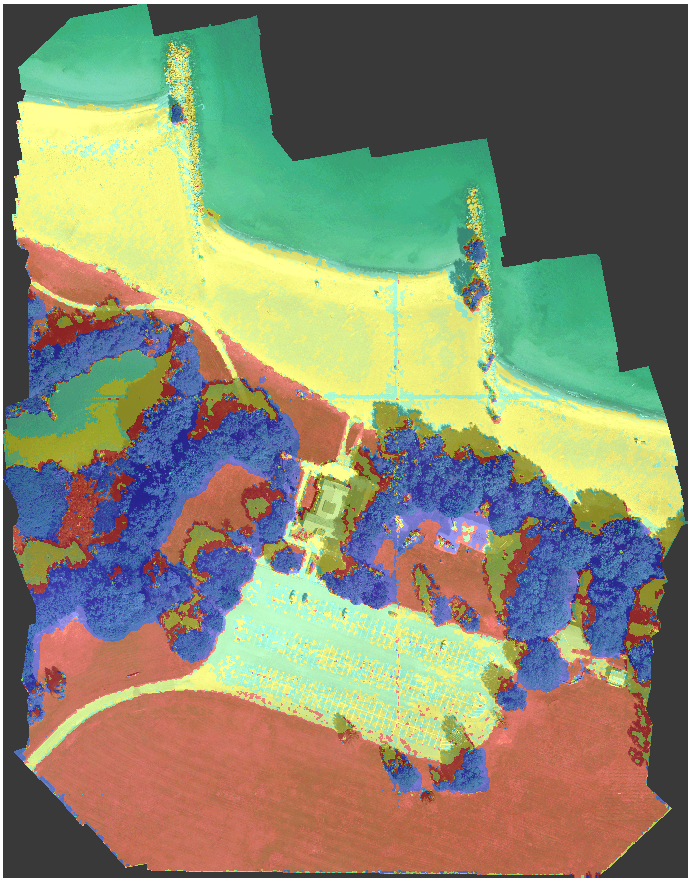}}}
	\caption{\small{Results of MSI segmentation by U-NET. (a) Result on the given LR image (x2). (b) Result on the I-TV$^{0}$ reconstruction.}}
	\label{fig:landcover2}
\end{figure}

In the second experiment we use the I-TV$^0$ model for SR to pre-process an image from the validation set of the HBSP dataset before giving it as an input to the U-Net \cite{kemker2017}. To do so, we consider a LR MSI acquisition of size $440 \times 350 \times 6$ and apply the SR model (with $L=2$) to each channel. For comparisons, we use the U-Net both on the given LR MSI and on the computed SR reconstruction. We report the results in Figure \ref{fig:landcover2}. Note, that differently from $k$-Means, U-NET does not require the user to specify the number of required classes.  
We observe that the quality of the U-Net segmentation is significantly improved when a pre-processing with SR I-TV$^{0}$ is made. When applied to the given LR image (see Figure \ref{fig:landcover2} (a)), the U-Net is in fact not capable to differentiate the group of trees (blue) from the grass (red). Increasing the resolution and promoting sparsity on the image gradient seems to be of great help for achieving more accurate results.

\subsubsection{Cell detection}
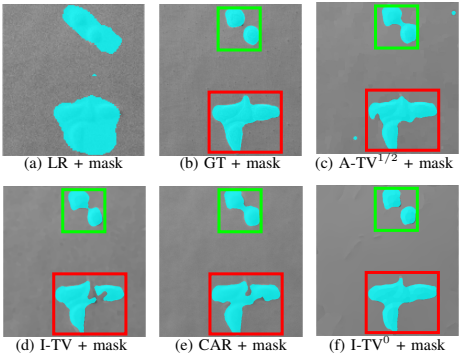
\begin{figure}
	\centering
	\scalebox{0.8}{
	\begin{tikzpicture}
	\begin{scope}[]
	\node [name=c] {\includegraphics[height=2.5
	cm]{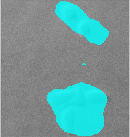}};
	\end{scope}
	\node[below of=c, align=center, node distance=1.4cm] {\small{\scalebox{0.85}{(a) LR + mask}}};
	\end{tikzpicture}\begin{tikzpicture}
	\begin{scope}[]
	\node [name=c] {\includegraphics[height=2.5
	cm]{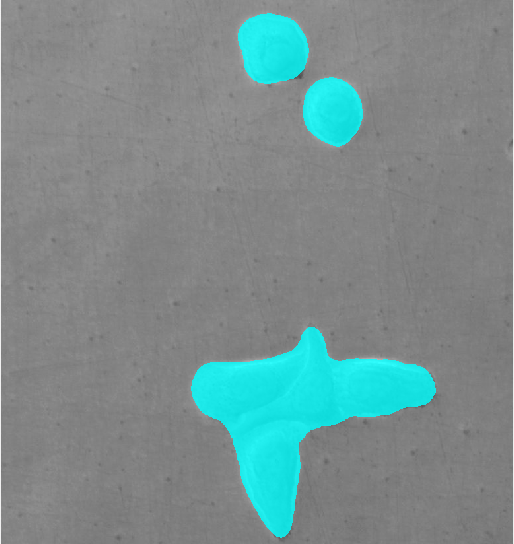}};
	\end{scope}
	\node[below of=c, align=center, node distance=1.4cm] {\small{\scalebox{0.85}{(b) GT + mask}}};
	\draw[green,line width=0.5mm] (-0.2,0.5) rectangle (0.5, 1.2);
	\draw[red,line width=0.5mm] (-0.35,-1.2) rectangle (0.85, -0.2);
	\end{tikzpicture}\begin{tikzpicture}
	\begin{scope}[]
	\node [name=c] {\includegraphics[height=2.5
	cm]{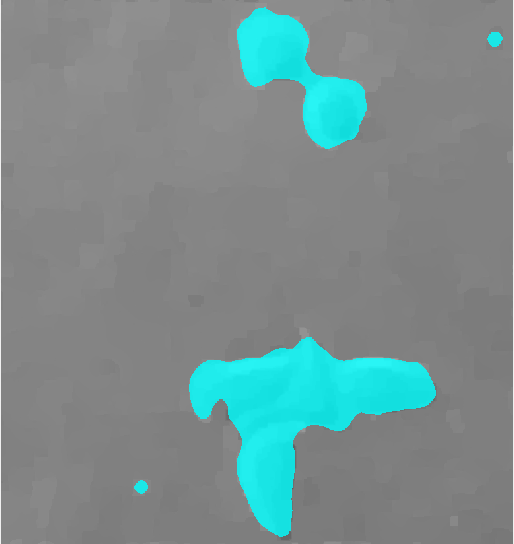}};
	\draw[red,line width=0.5mm] (-0.35,-1.2) rectangle (0.85, -0.2);
	\draw[green,line width=0.5mm] (-0.2,0.5) rectangle (0.5, 1.2);
	\end{scope}
	\node[below of=c, align=center, node distance=1.4cm] {\small{\scalebox{0.85}{(c) A-TV$^{1/2}$ + mask}} };
	\end{tikzpicture}	
	}
	
     \centering
     \scalebox{0.8}{
	\begin{tikzpicture}
	\begin{scope}[]
	\node [name=c] {\includegraphics[height=2.5
	cm]{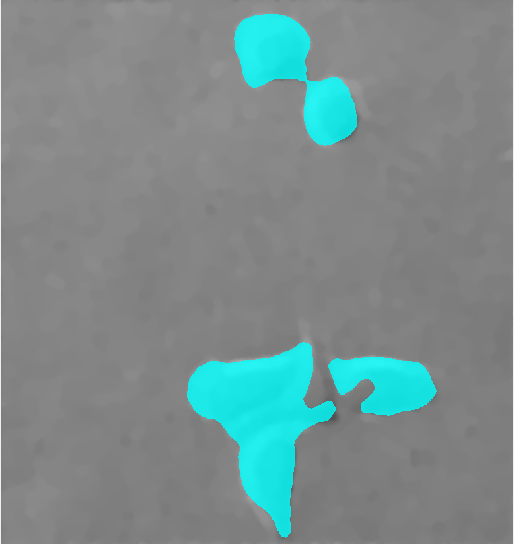}};
	\draw[red,line width=0.5mm] (-0.35,-1.2) rectangle (0.85, -0.2);
	\draw[green,line width=0.5mm] (-0.2,0.5) rectangle (0.5, 1.2);
	\end{scope}
	\node[below of=c, align=center, node distance=1.4cm] {\small{\scalebox{0.85}{(d) I-TV + mask}} };
	\end{tikzpicture}\begin{tikzpicture}
	\begin{scope}[]
	\node [name=c] {\includegraphics[height=2.5
	cm]{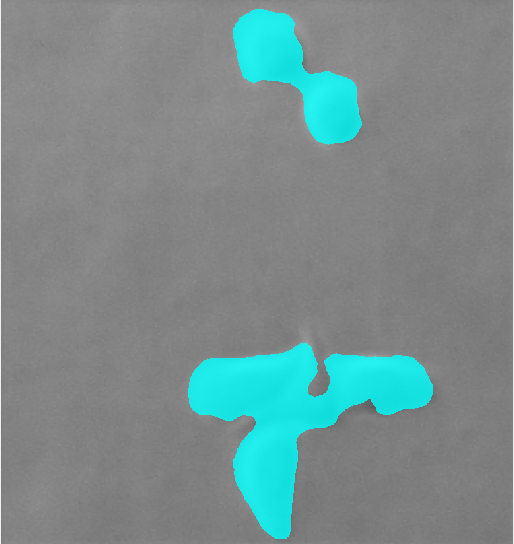}};
	\draw[green,line width=0.5mm] (-0.2,0.5) rectangle (0.5, 1.2);
	\draw[red,line width=0.5mm] (-0.35,-1.2) rectangle (0.85, -0.2);
	\end{scope}
	\node[below of=c, align=center, node distance=1.4cm] {\small{\scalebox{0.85}{(e) CAR + mask}}};
	\end{tikzpicture}\begin{tikzpicture}
	\begin{scope}[]
	\node [name=c] {\includegraphics[height=2.5
	cm]{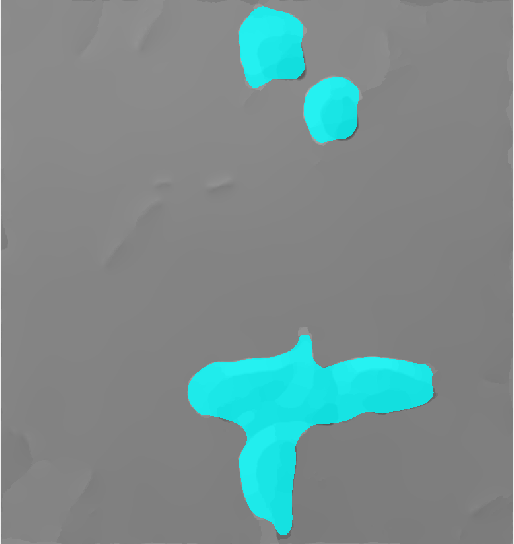}};
	\draw[green,line width=0.5mm] (-0.2,0.5) rectangle (0.5, 1.2);
	\draw[red,line width=0.5mm] (-0.35,-1.2) rectangle (0.85, -0.2);
	\end{scope}
	\node[below of=c, align=center, node distance=1.4cm] {\small{\scalebox{0.85}{(f) I-TV$^{0}$ + mask}}};
	\end{tikzpicture}
	}
	\caption{\small{Cell detection results. In (b)-(f) the green and red squares indicate two isolate cells and a cell cluster, respectively. Computed masks are coloured cyan.}}
	\label{fig:cell}

    \end{figure}

Standard light microscopes suffer from a limited resolving power which often causes blur artefacts and limits spatial resolution. In such conditions, the good performance of segmentation algorithms allowing for a precise location of isolated cells as well as cell clusters  is very challenging, despite their large use in biomedical applications where a fast segmentation is important for data analysis \cite{kothari2009automated}.
We test our $\ell^0$-gradient SR model on the light-microscope EVICAN data (Figure \ref{fig:cell} (a)) \cite{schwendy2020evican} for which the reference GT image in Figure \ref{fig:cell} (b)  has been found based on star-convexity shape prior \cite{gulshan2010geodesic}.
We apply the I-TV$^0$ model and its competitors on the LR acquisition obtained by \eqref{eq:model} from GT setting $L=4$, $\sigma_{H}=6$ and $\sigma_{\eta}=0.02$. For the different methods, the segmented regions are shown in Figure \ref{fig:cell} (c)-(e), while in Table \ref{tab:3} the PSNR, SSIM and Jaccard index values are reported. The $\ell^0$-gradient sparsity enforced by the I-TV$^0$ method allows for a better detection of the two isolated cells (green boxes) as well as the cell cluster (red boxes). However, when compared to I-TV, such simplification penalises more strongly image reconstruction metrics (PSNR and SSIM).



\begin{table}[h] 
\begin{center}
\caption{\small{Quantitative comparisons on cell image SR between different methods.}}
\scalebox{0.9}{
\begin{tabular}{c|ccc}\hline
\textbf{Method}  & \textbf{PSNR}  & \textbf{SSIM} & \textbf{Jaccard} \\ \hline 
I-TV   & \textcolor{red}{35.6891}   & \textcolor{red}{0.9198}   & 0.6855   \\ 
I-TV$^{1/2}$  & 35.2428                    & 0.9102                    &  0.8753      \\ 
I-TV$^{0}$ & 35.2863   & 0.9135   & \textcolor{red}{0.8778}    \\ 
CAR    & 35.1664   &  0.9044   & 0.8057    \\
\end{tabular} 
}
\label{tab:3}
\end{center}
\end{table}

	

\begin{figure*}
	\centering
	\begin{tikzpicture}
	
	\begin{scope}[spy using outlines={rectangle,red,magnification=3,size=4cm}]
	\begin{scope}[spy using outlines={rectangle,red,magnification=3,size=4cm}]
	\node {	\includegraphics[height=3cm]{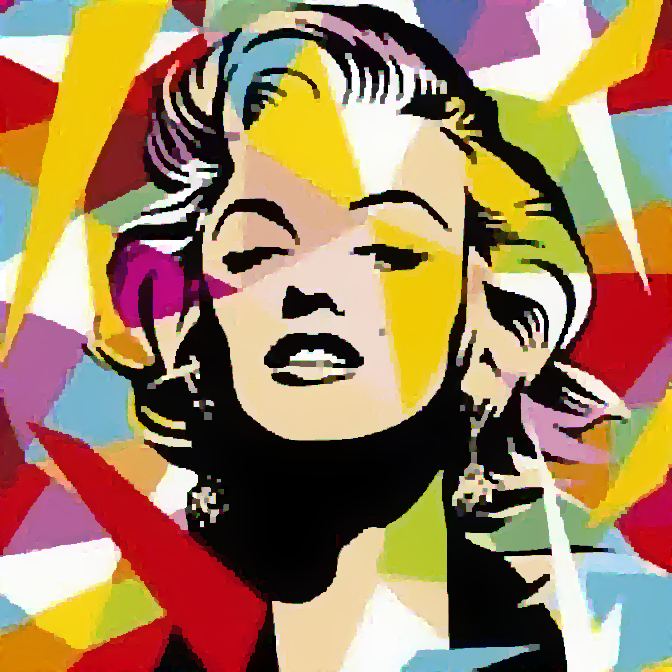}};
	\spy  [blue, width=1.5cm, height=1.5cm,magnification=5] on (-0.6,-0.7) in node [name=c6] at (10.35,0.78);
	\spy  [red, width=1.5cm, height=1.5cm] on (0.1,0) in node [name=c6] at (10.35,-0.78);
	\end{scope}
	\spy  [green!90, width=0.5cm, height=0.5cm,magnification=3] on (10.68,-0.73) in node [name=cz] at (10.8,-1.23);
	\end{scope}
	\node[below of=c6, align=center, node distance=1cm] {\scalebox{0.75}{A-TV$^{0}$}};
	
	\begin{scope}[spy using outlines={rectangle,red,magnification=3,size=4cm}]
	\begin{scope}[spy using outlines={rectangle,red,magnification=3,size=4cm}]
	\node {	\includegraphics[height=3cm]{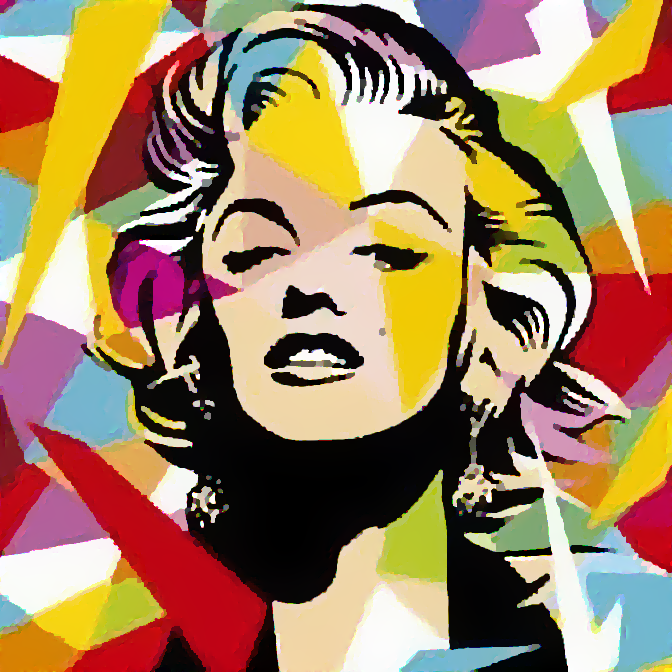}};
	\spy  [blue, width=1.5cm, height=1.5cm,magnification=5] on (-0.6,-0.7) in node [name=c5] at (11.95,0.78);
	\spy  [red, width=1.5cm, height=1.5cm] on (0.1,0) in node [name=c5] at (11.95,-0.78);
	\end{scope}
	\spy  [green!90, width=0.5cm, height=0.5cm,magnification=3] on (12.28,-0.73) in node [name=cz] at (12.4,-1.23);
	\end{scope}
	\node[below of=c5, align=center, node distance=1cm] {\scalebox{0.75}{I-TV$^{0}$}};
		
	\begin{scope}[spy using outlines={rectangle,red,magnification=3,size=4cm}]
	\begin{scope}[spy using outlines={rectangle,red,magnification=3,size=4cm}]
	\node {	\includegraphics[height=3cm]{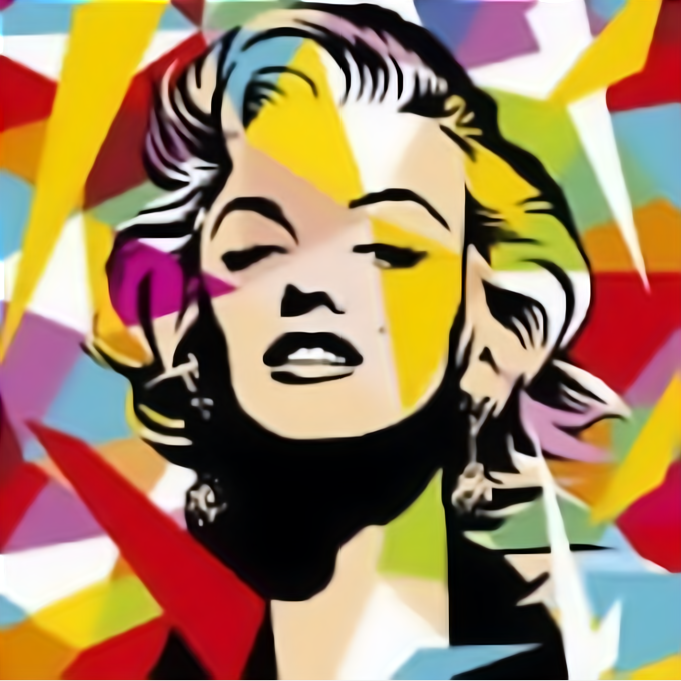}};
	\spy  [blue, width=1.5cm, height=1.5cm,magnification=5] on (-0.6,-0.67) in node [name=c4] at (8.75,0.78);
	\spy  [red, width=1.5cm, height=1.5cm] on (0.07,0.03) in node [name=c4] at (8.75,-0.78);
	\end{scope}
	\spy  [green!90, width=0.5cm, height=0.5cm,magnification=3] on (9.08,-0.73) in node [name=cz] at (9.2,-1.23);
	\end{scope}
	\node[below of=c4, align=center, node distance=1cm] {\scalebox{0.75}{IRCNN}};
	
	\begin{scope}[spy using outlines={rectangle,red,magnification=3,size=4cm}]
	\begin{scope}[spy using outlines={rectangle,red,magnification=3,size=4cm}]
	\node {	\includegraphics[height=3cm]{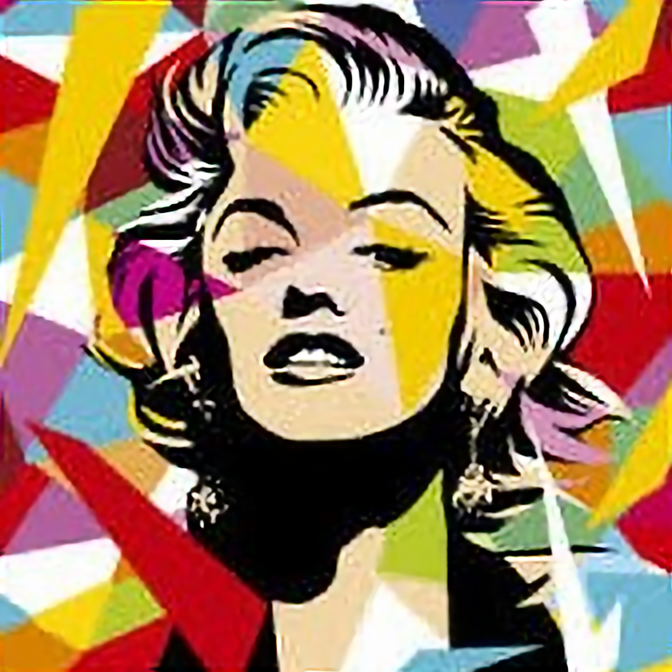}};
	\spy  [blue, width=1.5cm, height=1.5cm,magnification=5] on (-0.6,-0.67) in node [name=c3] at (7.15,0.78);
	\spy  [red, width=1.5cm, height=1.5cm] on (0.1,0.003) in node [name=c3] at (7.15,-0.78);
	\end{scope}
	\spy  [green!90, width=0.5cm, height=0.5cm,magnification=3] on (7.48,-0.73) in node [name=cz] at (7.6,-1.23);
	\end{scope}
	\node[below of=c3, align=center, node distance=1cm] {\scalebox{0.75}{CAR}};

	\begin{scope}[spy using outlines={rectangle,red,magnification=3,size=4cm}]
	\begin{scope}[spy using outlines={rectangle,red,magnification=3,size=4cm}]
	\node {	\includegraphics[height=3cm]{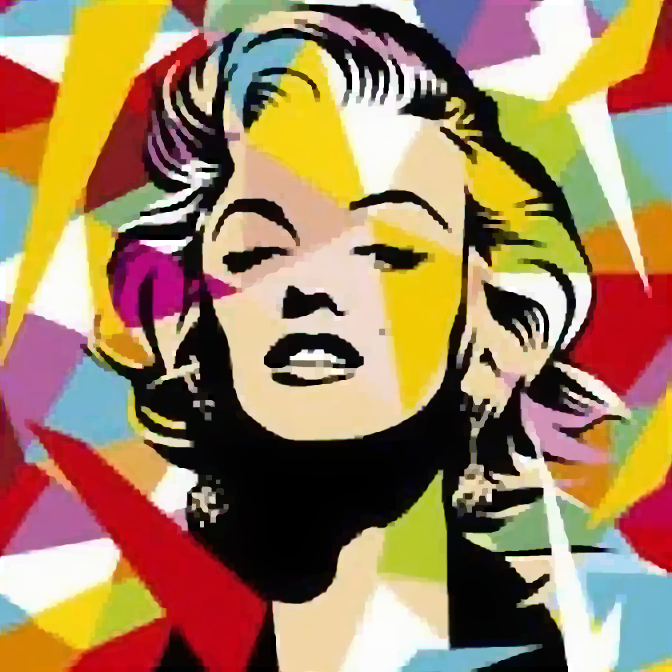}};
	\spy  [blue, width=1.5cm, height=1.5cm,magnification=5] on (-0.6,-0.7) in node [name=c2] at (5.55,0.78);
	\spy  [red, width=1.5cm, height=1.5cm] on (0.1,0) in node [name=c2] at (5.55,-0.78);
	\end{scope}
	\spy  [green!90, width=0.5cm, height=0.5cm,magnification=3] on (5.88,-0.73) in node [name=cz] at (6,-1.23);
	\end{scope}
	\node[below of=c2, align=center, node distance=1cm] {\scalebox{0.75}{c-TV}};
	
	\begin{scope}[spy using outlines={rectangle,red,magnification=3,size=4cm}]
	\begin{scope}[spy using outlines={rectangle,red,magnification=3,size=4cm}]
	\node {	\includegraphics[height=3cm]{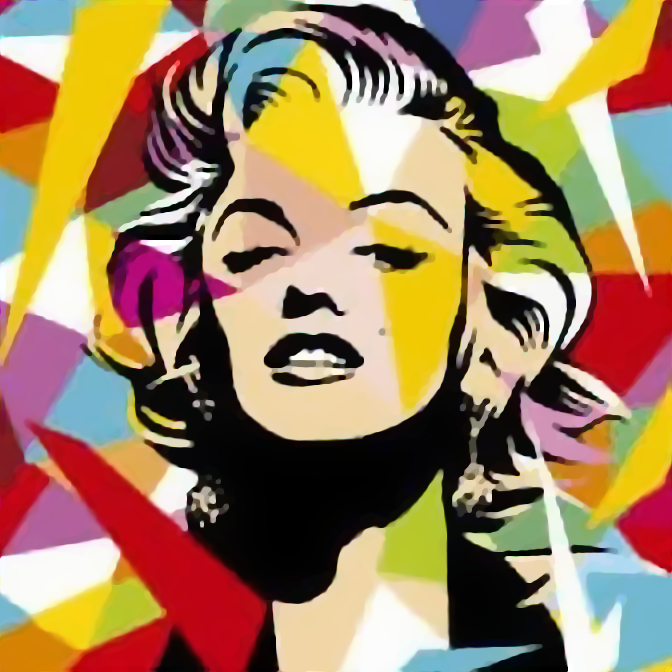}};
	\spy  [blue, width=1.5cm, height=1.5cm,magnification=5] on (-0.6,-0.7) in node [name=c1] at (3.95,0.78);
	\spy  [red, width=1.5cm, height=1.5cm] on (0.1,0) in node [name=c1] at (3.95,-0.78);
	\end{scope}
	\spy  [green!90, width=0.5cm, height=0.5cm,magnification=3] on (4.28,-0.73) in node [name=cz] at (4.4,-1.23);
	\end{scope}
	\node[below of=c1, align=center, node distance=1cm] {\scalebox{0.75}{I-TV}};
	
	\begin{scope}[spy using outlines={rectangle,red,size=4cm}]
	\begin{scope}[spy using outlines={rectangle,red,size=4cm}]
	\node[name=c]{	\includegraphics[height=3cm]{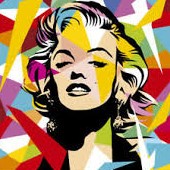}};
	\spy  [blue, width=1.5cm, height=1.5cm,magnification=5] on (-0.6,-0.67) in node [name=c1] at (2.35,0.78);
	\spy  [red, width=1.5cm, height=1.5cm,magnification=3] on (0.078,0.018) in node [name=c2] at (2.35,-0.78);
	\end{scope}
	\spy  [green!90, width=0.5cm, height=0.5cm,magnification=3] on (2.68,-0.73) in node [name=c3] at (2.8,-1.23);
	\end{scope}
	\node[below of=c2, align=center, node distance=1cm] {\scalebox{0.75}{LR (x4)}};
	\end{tikzpicture}
	
	\caption{\small{JPG artefact removal by means of different SR models. For I-TV, c-TV, A-TV$^{0}$ and I-TV$^{0}$ the regularisation parameters are chosen as $\mu$: 0.08, 0.05, 0.02, 0.02, respectively.}}
	\label{fig:MM}
\end{figure*}

\subsubsection{Compressed JPG images SR}  \label{sec:JPG}
In \cite{Xu11,Ono2017,Storath14}  $\ell^0$-gradient regularisation has been used for JPG compression artefact removal. Here, we consider a scenario where such task is performed along with a resolution improvement. To do so, we consider an RGB LR cartoon-type image of size $170 \times 170$ suffering from JPG compression artefacts and with small, not discernible details, and apply the gradient-sparse SR models. As no ground truth is available for this example, for all models we empirically select the parameters producing the best visual output. 

 In Figure \ref{fig:MM} we report two close-ups of the computed SR reconstructions marked by blue and red boxes. The blue box highlights small details which are poorly discernible in the LR image, while the red box considers a patch of the face with some blunt edges and a small (but meaningful!) face mole (see green box). We see that both A-TV$^{0}$ and I-TV$^{0}$ reconstructions are sharper and more cartoonised than the ones obtained by the other models.  Furthermore, the well-known I-TV and c-TV loss of contrast reconstruction artefact makes small details hardly discernible. Due to the high-level of compression artefacts, we remark that both IRCNN and CAR results are very blurred.

\section{Conclusions}

We considered a variational model with $\ell^0$ gradient-sparsity-promoting regularisation combined with a quadratic data fidelity for single-image super-resolution of images corrupted by blur and Gaussian noise. The use of non-convex $\ell^0$ jump-sparse regularisations has been originally proposed in \cite{Storath14} in the context of general 1D inverse problems and subsequently applied in \cite{Storath15,Storath17} to image segmentation and reconstruction problems. To overcome the computational limitations required by the use of ADMM splitting strategies considered in these works, we propose a novel ADMM algorithm allowing for the efficient solution of its subproblems by means of direct hard-thresholding or standard CG solvers. For the proposed scheme we prove fixed-point convergence results assuming specific growth conditions on the sequence of penalty parameters. We validate our model on synthetic data and test it on real-world examples where gradient-sparse super-resolved outputs are required in view of an accurate recognition/classification step (such as QR code recognition \cite{Kato2011}, cell detection and land-cover classification \cite{cihlar2000}).  By numerous comparisons with convex and non-convex variational approaches, and with state-of-the-art deep learning methods \cite{zhang2017,sun2020learned}, we show that the proposed approach significantly improves classification precision, while limiting at the same times smoothing and loss-of-contrast artefacts in comparison with classical convex regularisations.

Further work should address the use of analogous regularisations and algorithms for the joint modelling of SR and segmentation problems via, e.g., Mumford-Shah functionals \cite{Storath15}. Furthermore, the extension of the convergence results to other gradient discretisations and to less restrictive growth conditions for the sequence of penalty parameters is envisaged. 



%

\appendices

\section{Convergence analysis} \label{sec:convergence}

We report here a complete convergence proof of Theorem \ref{th:conv_ADMManiso} and a sketch of the proof of Theorem \ref{th:conv_ADMMiso}, which is based on similar arguments.

\subsection{Proof of Theorem \ref{th:conv_ADMManiso}}

\begin{proof} 
We consider the ADMM sequences $(\ubold^{k}),(\tbold^{k}),(\sbold^{k})$, defined in \eqref{eq:ADMM_aniso_1}-\eqref{eq:ADMM_aniso_3}. We want to show that there exists $\ubold^*$ such that:
\begin{align}
\ubold^{k} \rightarrow \ubold^{*}, \quad \tbold^{k} \rightarrow\Dbold_{h}\ubold^{*}\nonumber, \quad \sbold^{k} \rightarrow  \Dbold_{v}\ubold^{*}. 
\end{align}
To shorten the proof, we remark that everything proved for the sequences $(\tbold^{k})$,$(\beta_{t}^{k})$, $(\boldsymbol{\lambda}_{t}^{k})$ and $(\Dbold_{h}\ubold^{k})$ can be deduced for $(\sbold^{k})$, $(\beta_{s}^{k})$, $(\boldsymbol{\lambda}_{s}^{k})$ and $(\Dbold_{v}\ubold^{k})$ in the same way.

We start defining the following functionals:
\begin{align*}
&G^{h}_{k}(\tbold):= \mu \lVert \tbold \rVert_{0} + \dfrac{\beta_{t}^{k}}{2} \lVert \tbold - (\Dbold_{h}\ubold^{k}+\dfrac{\boldsymbol{\lambda}_{t}^{k}}{\beta_{t}^{k}})\rVert_{2}^{2}, \\
& F_{k}(\ubold):= \dfrac{1}{2}\lVert \Sbold \Hbold \ubold - \gbold \rVert_{2}^{2}  + \dfrac{\beta_{t}^{k}}{2}\lVert \Dbold_{h}\ubold - (\tbold^{k+1} - \dfrac{\boldsymbol{\lambda}_{t}^{k}}{\beta_{t}^{k}})\rVert^{2}_{2} + \nonumber \\  & + \dfrac{\beta_{s}^{k}}{2}\lVert \Dbold_{v}\ubold - (\sbold^{k+1} - \dfrac{\boldsymbol{\lambda}_{s}^{k}}{\beta_{s}^{k}})\rVert^{2}_{2}.
\end{align*}
\paragraph*{Step 1}
There holds: 
\begin{align}
&\lVert \tbold^{k+1} -\Dbold_{h}\ubold^{k} - \dfrac{\boldsymbol{\lambda}_{t}^{k}}{\beta_{t}^{k}} \rVert_{2} \leq \sqrt{\dfrac{2 \mu N}{\beta_{t}^{k}}}. \label{proof:step1_limitato_t} 
\end{align}
This inequality can  be trivially shown by the minimality of $\tbold^{k+1}$ in \eqref{eq:ADMM_aniso_1} which entails   $G^{h}_{k}(\tbold^{k+1}) \leq  G^{h}_{k}(\Dbold_{h}\ubold^{k} + \frac{\boldsymbol{\lambda}_{t}^{k}}{\beta_{t}^{k}})$, therefore we get:
\begin{align*}
& \mu \lVert \tbold^{k+1}\rVert_{0} + \dfrac{\beta_{t}^{k}}{2} \lVert \tbold^{k+1} - (\Dbold_{h}\ubold^{k}+\dfrac{\boldsymbol{\lambda}_{t}^{k}}{\beta_{t}^{k}})\rVert_{2}^{2} \notag\\
& \leq \mu \lVert \Dbold_{h}\ubold^{k} + \frac{\boldsymbol{\lambda}_{t}^{k}}{\beta_{t}^{k}} \rVert_{0} \leq \mu N,
\end{align*}
by definition of $\lVert \cdot \rVert_0$,where we recall $N$ is the dimension of the vector $\ubold^{k}$. By neglecting the first term on the Left Hand Side (LHS)  of the above inequality, we deduce \eqref{proof:step1_limitato_t}. 

\paragraph*{Step 2}
From the minimality of $\ubold^{k+1}$ in \eqref{eq:ADMM_aniso_3} we have: $F_{k}(\ubold^{k+1}) \leq F_{k}(\ubold^{k})$ for every $k$. By definition of $F_{k}$ and applying \eqref{proof:step1_limitato_t} and its analogous related to the sequences $(\sbold^{k})$, $(\beta_{s}^{k})$, $(\boldsymbol{\lambda}_{s}^{k})$ and $(\Dbold_{v}\ubold^{k})$, we deduce: 
\begin{align}
    & \dfrac{1}{2}\lVert \Sbold \Hbold \ubold^{k+1} - \gbold \rVert_{2}^{2}  + \dfrac{\beta_{t}^{k}}{2} \lVert \Dbold_{h} \ubold^{k+1} - \tbold^{k+1} + \dfrac{\boldsymbol{\lambda}_{t}^{k}}{\beta_{t}^{k}} \rVert_{2}^{2}  \label{proof:riferimento} \\
    & + \dfrac{\beta_{s}^{k}}{2} \lVert \Dbold_{v} \ubold^{k+1} - \sbold^{k+1} + \dfrac{\boldsymbol{\lambda}_{s}^{k}}{\beta_{s}^{k}}  \rVert_{2}^{2} \leq  
   \dfrac{1}{2}\lVert \Sbold \Hbold \ubold^{k} - \gbold \rVert_{2}^{2} + 2\mu N. \nonumber
\end{align}
Since the all the terms on the LHS of \eqref{proof:riferimento} are nonnegative, the following inequality holds: 
\begin{align}
    & \dfrac{1}{2} \lVert \Sbold \Hbold \ubold^{k+1} -\gbold \rVert_{2}^{2} \leq \dfrac{1}{2} \lVert \Sbold \Hbold \ubold^{k} -\gbold \rVert_{2}^{2} + 2 \mu N \leq \ldots \\
    & \leq  \dfrac{1}{2} \lVert \Sbold \Hbold \ubold^{0} - \gbold \rVert_{2}^{2} +  2 \mu N k \nonumber
\end{align}
From \eqref{proof:riferimento} and by the sub-additivity property of the square root we can also derive the following inequality:
\begin{align}\label{ineq:series_convergent2}
    &\lVert \Dbold_{h}\ubold^{k+1} - \tbold^{k+1} + \dfrac{\boldsymbol{\lambda}_{t}^{k}}{\beta_{t}^{k}}\rVert_{2}    \leq \sqrt{\frac{1}{{\beta_{t}^{k}}}}\lVert \Sbold \Hbold \ubold^{0} - \gbold \rVert_{2} + \sqrt{4\mu N \frac{k}{\beta_{t}^{k}}}
\end{align}

\paragraph*{Step 3}
We show that the sequences $\Dbold_{h}\ubold^{k}$ and $\Dbold_{v}\ubold^{k}$ are Cauchy sequences, hence they converge. We prove this for $\Dbold_{h}\ubold^{k}$, the proof for $\Dbold_{v}\ubold^{k}$ is identical.
\begin{align*}
    & \lVert \Dbold_{h} \ubold^{k+1} - \Dbold_{h} \ubold^{k} \rVert_{2} \leq \nonumber \\
    &\leq \lVert \Dbold_{h}\ubold^{k+1} - \tbold^{k+1} +  \dfrac{\boldsymbol{\lambda}_{t}^{k}}{\beta_{t}^{k}}  \rVert_{2}  +   \lVert \Dbold_{h}\ubold^{k} - \tbold^{k+1} +  \dfrac{\boldsymbol{\lambda}_{t}^{k}}{\beta_{t}^{k}}  \rVert_{2} . 
\end{align*}
By assumption \ref{cond:1_ADMManiso} applied on the RHS of \eqref{ineq:series_convergent2} we deduce:
\begin{align} 
    &\lVert \Dbold_{h}\ubold^{k+1} - \tbold^{k+1} + \dfrac{\boldsymbol{\lambda}_{t}^{k}}{\beta_{t}^{k}}\rVert_{2} \rightarrow 0, \label{prof:bound1}
\end{align}
which, combined with \eqref{proof:step1_limitato_t} and \eqref{ineq:series_convergent2} entails that $\Dbold_{h}\ubold^{k}$ is a Cauchy sequence. Hence  it converges to a point $\tbold^{*}$. Similarly, $\Dbold_{v}\sbold^{k}$  converges to a point $\sbold^{*}$.


\paragraph*{Step 4}
We prove now the convergence of the sequences $\tbold^{k}$ and $\Dbold_{h} \ubold^{k}$.
By writing \eqref{eq:ADMM_aniso_4} as: 
\begin{align}
 &\frac{\boldsymbol{\lambda}_{t}^{k+1}}{\beta_{t}^{k}} = \Dbold_{h} \ubold^{k+1} - \tbold^{k+1} + \dfrac{\boldsymbol{\lambda}_{t}^{k}}{\beta_{t}^{k}}, \label{eq:proof_dual1}
\end{align}
and  from \eqref{prof:bound1} we deduce that $\frac{\lVert \boldsymbol{\lambda}_{t}^{k+1} \rVert_{2}}{\sqrt{\beta_{t}^{k}}} \rightarrow 0 $. By monotonicity of the $(\beta^k_t)$ we then deduce that  $\frac{\lVert \boldsymbol{\lambda}_{t}^{k} \rVert_{2}}{\sqrt{\beta_{t}^{k}}} \rightarrow 0 $. Hence:
\begin{align*}
& \lVert \Dbold_{h}\ubold^{k+1} - \tbold^{k+1} \rVert_{2} \leq  \dfrac{\lVert \boldsymbol{\lambda}_{t}^{k+1}\rVert_{2} + \lVert \boldsymbol{\lambda}_{t}^{k}\rVert_{2}}  {\sqrt{\beta_{t}^{k}}}, 
\end{align*}
where both quantities on the RHS tend to $0$ as $k\to\infty$. Therefore, by the uniqueness of the limit, $\tbold^{k} \longrightarrow \tbold^{*}$ and $\Dbold_{h}\ubold^{k} \longrightarrow \tbold^{*}$.  

\paragraph*{Step 5}
We can now prove convergence of the sequence $(\ubold^{k})$ . For simplicity, let us define the quantities $\Abold:=\Sbold\Hbold$ and  $\Mbold_{k}:=\dfrac{1}{\beta_t^{k}} \Abold^{T}\Abold + \Dbold_{h}^{T}\Dbold_{h} + \dfrac{\beta_{s}^{k}}{\beta_{t}^{k}}\Dbold_{v}^{T}\Dbold_{v}$, for every $k$. By \ref{cond:2_ADMManiso}, we observe that the matrix $\Mbold_k$ is invertible for all $k$ and that the optimality condition of \eqref{eq:ADMM_aniso_3} reads:
\begin{equation}
\Mbold_{k}\ubold^{k}= \Dbold_{h}^{T} (\tbold^{k+1} - \dfrac{\boldsymbol{\lambda}_{t}^{k}}{\beta^{k}}) + \dfrac{\beta_{s}^{k}}{\beta_{t}^{k}}\Dbold_{v}^{T}(\sbold^{k+1} - \dfrac{\boldsymbol{\lambda}_{s}^{k}}{\beta^{k}}) + \dfrac{1}{\beta_{t}^{k}}\Abold^{T}\gbold. \nonumber
\end{equation}
Since $\tbold^{k+1} \to \tbold^{*}$, $\sbold^{k+1} \to \sbold^{*}$,  $\frac{\boldsymbol{\lambda}_{t}^{k}}{\beta_{t}^{k}} \to 0$, $\frac{\boldsymbol{\lambda}_{s}^{k}}{\beta_{s}^{k}} \to 0$,  and by Assumptions \ref{cond:1_ADMManiso} and \ref{cond:2_ADMManiso},  we have that   $\frac{1}{\beta_{t}^{k}}\Abold^{T}\gbold \to \mathbf{0}$ so that the RHS converges pointwise to $\zbold^{*}=\Dbold_{h}^{T}t^{*} + c \Dbold_{v}^{T}s^{*} $. Additionally, the sequence $\Mbold_{k}^{-1}$ converges pointwise to $\Mbold^{*}$. We thus have that $\ubold^k=\Mbold_{k}^{-1}\Mbold_{k}\ubold^{k} \to \Mbold^{*}\zbold^{*}:=\ubold^{*}$. \\
We now want to show that $\tbold^*=\Dbold_h \ubold^*$ and, similarly, that $\sbold^*=\Dbold_v \ubold^*$. We show the details only for the former case. By the triangle inequality we get:
\begin{align}
\lVert \tbold^* -  \Dbold_h \ubold^*\rVert_2  \leq & \lVert \tbold^* -  \Dbold_h \ubold^k\rVert_2 + \lVert \Dbold_h \ubold^k - \Dbold_h \ubold^* \rVert_2 \nonumber \\
& \leq \lVert \tbold^* -  \Dbold_h \ubold^k\rVert_2 + \lVert \Dbold_h \rVert_2 \lVert \ubold^k - \ubold^* \rVert_2, \nonumber
\end{align}
where both terms tend to $\mathbf{0}$ since $\Dbold_h \ubold^k \to \tbold^* $ and $\ubold^k \to \ubold^*$. 
\end{proof}



\subsection{Proof of Theorem \eqref{th:conv_ADMMiso}}

\begin{proof}
The proof of Theorem \eqref{th:conv_ADMMiso} follows the same steps as the previous one. The only main difference in it is the definiton of $\Mbold_{k}$, which reads in this case:
\begin{align}
    \Mbold_{k}:= \dfrac{1}{\beta^{k}} \Abold^{T}\Abold + \Dbold^{T}\Dbold =
    \dfrac{1}{\beta^{k}} \Abold^{T}\Abold+
    \Dbold_{h}^{T}\Dbold_{h} + \Dbold_{v}^{T}\Dbold_{v}. \nonumber
\end{align}
By proceeding similarly as above the conclusion holds.
\end{proof}


\section*{Acknowledgments}

LC and PC acknowledge the support received by the Academy "Complex Systems" of the JEDI IDEX of the Université Côte d'Azur.
ELP and PC acknowledge the support received by the INDAM-GNCS (Research projects 2020).

\ifCLASSOPTIONcaptionsoff
  \newpage
\fi



%


\vspace{-1cm}




%



\begin{IEEEbiography}
[{\includegraphics[width=1in,height=1in,clip=true]{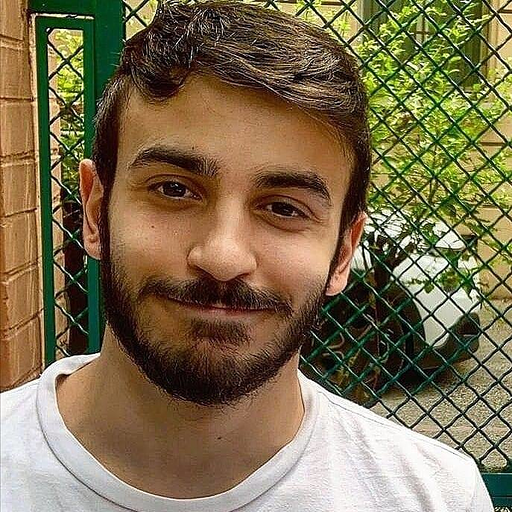}}]{Pasquale Cascarano}
received his master degree in Mathematics in 2018 at University of Bologna. He is currently a Ph.D.\ student in Applied Mathematics at the University of Bologna (IT). His research focuses on variational and deep learning methods for imaging inverse problems.
\end{IEEEbiography}
\vspace{-1.5cm}
\begin{IEEEbiography}
[{\includegraphics[width=1in,height=1.3in,trim=1.5cm 7cm 2.5cm 0cm,clip=true]{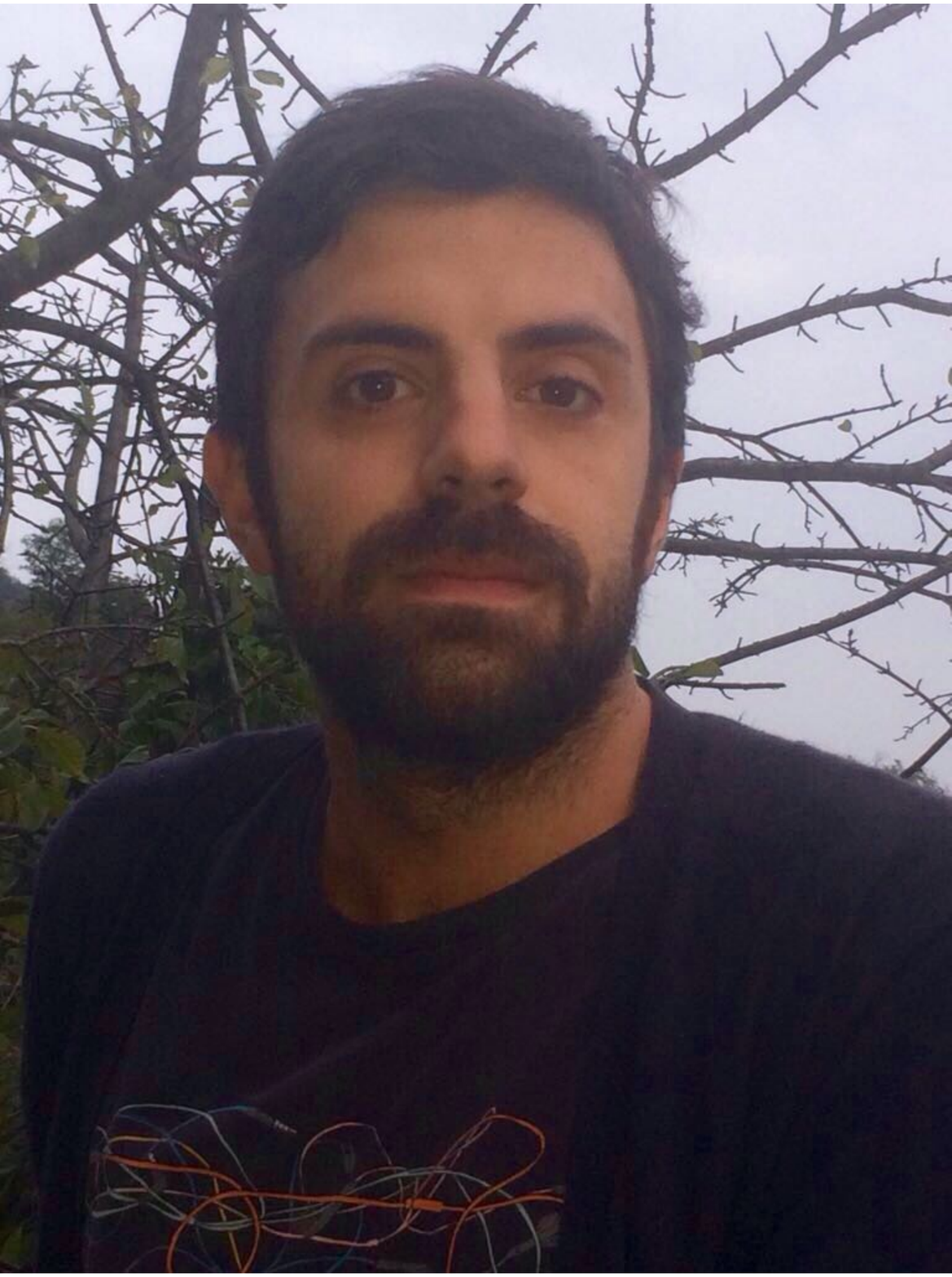}}]{Luca Calatroni}
completed his Ph.D.\ in Applied Mathematics in 2015 as part of the Cambridge Image Analysis research group (UK). He was then a Marie Skłowdoska-Curie research fellow at the University of Genova (Italy) and \emph{Lecteur Hadamard} FMJH fellow at the \'Ecole Polytechnique (France). From October 2019, he is permanent CNRS researcher at the I3S laboratory in Sophia Antipolis (France) within the Morpheme research group. His research focuses on variational methods and non-smooth optimisation algorithms for imaging and vision.
\end{IEEEbiography}
\vspace{-1.2cm}
\begin{IEEEbiography}
[{\includegraphics[width=0.8in,height=1.3in,clip=true]{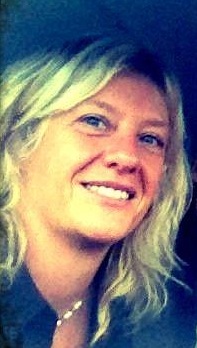}}]{Elena Loli Piccolomini} is Professor of Numerical Analysis at the University of Bologna. Her research topics are regularisation methods for inverse problems in imaging, with particular focus on medical imaging and tomographic image reconstruction.
\end{IEEEbiography}




\end{document}